\documentclass[11pt]{article}
\usepackage[T1]{fontenc}
\usepackage{lmodern}
\usepackage[margin=1in]{geometry}
\usepackage{microtype}
\usepackage{amsmath,amssymb,amsthm}
\usepackage{mathtools}
\usepackage{enumitem}
\usepackage{booktabs}
\usepackage{tabularx}
\usepackage{float}
\floatstyle{ruled}
\newfloat{algorithm}{tbp}{loa}
\floatname{algorithm}{Algorithm}
\usepackage{xcolor}
\usepackage{tikz}
\usepackage[colorlinks=true,linkcolor=blue!50!black,citecolor=blue!50!black,urlcolor=blue!50!black]{hyperref}
\newtheorem{theorem}{Theorem}[section]
\newtheorem{lemma}[theorem]{Lemma}
\newtheorem{corollary}[theorem]{Corollary}
\newtheorem{proposition}[theorem]{Proposition}
\DeclareMathOperator{\diam}{diam}
\DeclareMathOperator{\dist}{dist}
\DeclareMathOperator{\cost}{cost}
\DeclareMathOperator{\tw}{tw}
\newcommand{\OPT}{\mathcal O}
\newcommand{\eps}{\varepsilon}
\newcommand{\DP}{\mathrm{DP}}
\setlist[itemize]{topsep=4pt,itemsep=3pt}
\setlist[enumerate]{topsep=4pt,itemsep=4pt}
\title{Exact Sum-of-Radii Clustering under Any Constant Stability}
\author{Shyam Ravichandran \qquad Lev Reyzin\\[3pt]
  {\small Department of Mathematics, Statistics, and Computer Science}\\
  {\small University of Illinois Chicago}\\
  {\small\texttt{\{sravi42,lreyzin\}@uic.edu}}}
\date{}
\hypersetup{pdftitle={Exact Sum-of-Radii Clustering under Any Constant Stability},
  pdfauthor={Shyam Ravichandran, Lev Reyzin}}
\begin{document}
\maketitle

\begin{abstract}
We establish a separation between the stability thresholds of
Minimum Sum-of-Radii (MSR) clustering and two closely related
objectives. Known hardness results show that, on general metrics,
exact recovery for Minimum
Sum-of-Diameters and symmetric $k$-center is hard for every fixed
$1<\gamma<2$, unless $\mathrm{NP}=\mathrm{RP}$. We show that MSR
admits deterministic polynomial-time exact recovery for every fixed
$\gamma>1$, extending Gao and Reyzin's result for $\gamma\ge2$.
All three results use the same Bilu--Linial stability model: the
optimal partition remains uniquely optimal when each pairwise
distance is multiplied by any factor in $[1,\gamma]$, without
requiring the perturbed distances to satisfy the triangle inequality.
The number of clusters is part of the input. For $1<\gamma<2$,
our algorithm runs in $n^{O((\gamma-1)^{-5})}$ time.

The main structural insight is that a cheap merge in one part of the
instance would allow a profitable split elsewhere under a suitable
perturbation, contradicting stability. The resulting exchange
inequality controls local cluster density, which gives linear growth
and hence bounded treewidth in local interaction graphs. We use
random partitions to obtain these regions, solve them by dynamic
programming, and then
derandomize the construction. For rational input and fixed rational
$\gamma>1$, recognizing stability in this model is coNP-complete.
Verifying that a supplied partition remains uniquely optimal under
all allowed perturbations is coNP-complete for both arbitrary
symmetric and metric perturbations.
\end{abstract}

\section{Introduction}
\label{sec:intro}

Minimum Sum-of-Radii (MSR), Minimum Sum-of-Diameters (MSD), and
$k$-center partition a finite metric space into $k$ clusters. Their
objectives are, respectively, the sum of cluster radii, the sum of
cluster diameters, and the maximum cluster radius. MSR allows a
larger radius in one cluster to be offset by smaller radii elsewhere
and is motivated by coverage problems such as placing base stations
to serve demand points~\cite{Gibson10}.

These closely related objectives have different thresholds for
recovery under Bilu--Linial stability. On general metrics, when
perturbations may be arbitrary symmetric distance increases,
MSD~\cite[Theorem~IV.2]{GaoReyzin25} and symmetric
$k$-center~\cite[Theorem~10]{BalcanHW16} remain hard for every fixed
$1<\gamma<2$, unless $\mathrm{NP}=\mathrm{RP}$. In the same model,
we recover the unique MSR optimum in polynomial time for every fixed
$\gamma>1$.

Formally, for a nonempty set $C\subseteq P$, let
\[
 r_d(C)=\min_{c\in C}\max_{x\in C}d(c,x),\qquad
 \cost_d(\mathcal P)=\sum_{C\in\mathcal P}r_d(C).
\]
We require exactly $k$ nonempty clusters and choose each center from
its own cluster. Singletons have radius zero. We write $n=|P|$ and
omit the subscript when distances are measured by $d$.

A \emph{$\gamma$-perturbation} of $d$ is a symmetric function $d'$
satisfying $d(x,y)\le d'(x,y)\le\gamma d(x,y)$ for every pair of
points. We do not require $d'$ to satisfy the triangle inequality.
An instance is \emph{$\gamma$-stable} if a partition $\OPT$ is uniquely
optimal under every such perturbation, including $d$ itself. The
centers need not be unique or remain unchanged. This is the
perturbation-stability model of Bilu and Linial~\cite{BiluLinial12}.

Gao and Reyzin~\cite{GaoReyzin25} showed that single-linkage recovers
the optimum of every $2$-stable MSR instance. We give a deterministic
algorithm for the remaining range $1<\gamma<2$. By monotonicity of
stability, our algorithm also handles $\gamma\ge2$ by using parameter $3/2$.
Let $\eps=\gamma-1$.

\begin{samepage}
\begin{theorem}
\label{thm:main}
For every fixed $\gamma>1$, there is a deterministic polynomial-time
algorithm that recovers the unique optimum of every $\gamma$-stable
MSR instance under arbitrary symmetric perturbations. For
$1<\gamma<2$, its running time is
\[
 n^{O((\gamma-1)^{-5})}.
\]
On every finite metric input with $1\le k\le n$, it halts and returns
a feasible partition into exactly $k$ nonempty clusters.
\end{theorem}
\end{samepage}

Theorem~\ref{thm:main} separates the stability threshold for MSR from
the threshold $2$ for MSD~\cite[Theorem~III.2]{GaoReyzin25} and
$k$-center~\cite[Theorem~4]{BalcanHW16}. The recovery algorithm rests
on a structural consequence of stability: local cluster density is
bounded strongly enough to yield bounded-treewidth interaction graphs.
The key exchange argument uses the fact that fixing the number of clusters
links otherwise distant parts of the instance. Merging two optimal
clusters $D,E$ frees one place in the cluster count, allowing a third
cluster $C$ to be split elsewhere. For every nonsingleton $C$, we
construct a perturbation that increases its radius by $\eps r(C)$,
while splitting off a suitable point removes that increase.
The merge is unaffected by the perturbation. If it costs at most
$\eps r(C)$ extra, the merge and split together give a different
partition of no greater perturbed cost. Stability therefore forces
\[
 r(D\cup E)>r(D)+r(E)+\eps r(C).
\]
More generally, merging several clusters must cost more than the
savings from the corresponding number of splits elsewhere. The
clusters being split need not be near those being merged.

This exchange principle controls local cluster density at the radii
used by our algorithm. Too many small clusters meeting a ball could
be merged cheaply, allowing profitable splits of larger clusters
elsewhere and contradicting stability. Large clusters satisfy a
complementary bound from ordinary optimality: the sum of the radii
of optimal clusters contained in a ball cannot exceed the ball's
radius. Otherwise, one ball would cover them at smaller cost, and
we could convert this cheaper cover back to exactly $k$ clusters.
Together, these bounds limit how many clusters can meet a ball
without any assumption on the dimension of the metric.

The algorithmic consequence is the progression
\[
 \begin{gathered}\text{merge-and-split}\\\text{exchange}\end{gathered}
 \ \Longrightarrow\
 \begin{gathered}\text{local cluster}\\\text{density}\end{gathered}
 \ \Longrightarrow\
 \begin{gathered}\text{linear graph}\\\text{growth}\end{gathered}
 \ \Longrightarrow\
 \begin{gathered}\text{bounded}\\\text{treewidth}\end{gathered}.
\]
Within each region of sufficiently small diameter, choose points
separated by more than twice the largest optimal radius and join
points within a fixed multiple of that radius. Distinct vertices
belong to distinct optimal clusters. The density bound then limits
the number of vertices reachable in $j$ graph steps to $O_\eps(j)$
for $j\ge1$.
A theorem of Campbell et al.~\cite{Campbell23} converts this linear
growth into treewidth bounded solely by $\eps$, permitting exact
dynamic programming even when a region contains many clusters.
The exponent comes from $O(\eps^{-4})$ vertices per bag and
$n^{O(1/\eps)}$ candidate choices per vertex; enumeration itself
costs only $n^{O(\eps^{-2})}$.
These bounds establish polynomial-time tractability; they are too
large for a practical algorithm.

\begin{figure}[tbp]
\centering
\begin{minipage}[c]{0.43\linewidth}
\centering
\begin{tikzpicture}[x=0.94cm,y=0.94cm,font=\small]
  \path[use as bounding box] (-1.72,-1.08) rectangle (2.93,1.08);
  \draw[dashed,rounded corners=7pt,gray] (-1.62,-1.03) rectangle (1.97,1.03);
  \draw (-1.25,0.68) -- node[above,sloped] {$1$} (0,0);
  \draw (-1.25,-0.68) -- node[below,sloped] {$1$} (0,0);
  \draw (0,0) -- node[above] {$1$} (1.55,0);
  \draw (1.55,0) -- node[above,fill=white,inner sep=1pt] {$1/2$} (2.55,0);
  \foreach \x/\y/\lab/\pos in {0/0/o/below,1.55/0/a/below,2.55/0/p/below,-1.25/0.68/b/above,-1.25/-0.68/c/below}
    \fill (\x,\y) circle (0.055) node[\pos,inner sep=3pt] {$\lab$};
\end{tikzpicture}
\end{minipage}\hfill
\begin{minipage}[c]{0.54\linewidth}
\caption{A stable instance on which single-linkage fails.
The dashed enclosure is the radius-$1$ optimal cluster;
$p$ is a singleton. Single-linkage first merges $a,p$ along
the unique length-$1/2$ edge.}
\label{fig:single-linkage}
\end{minipage}
\end{figure}
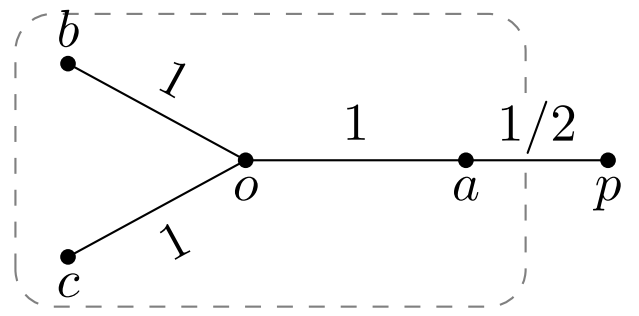

The need to go beyond single-linkage already appears on five points
(Figure~\ref{fig:single-linkage}).
Take the shortest-path metric of a tree with a center $o$, three
leaves $a,b,c$ joined to $o$ by unit edges, and a point $p$ joined
to $a$ by an edge of length $1/2$. For $k=2$, the optimum is
$\{\{o,a,b,c\},\{p\}\}$, of cost $1$; every other partition
costs at least $3/2$. Under any $\gamma$-perturbation, its cost
is at most $\gamma$, while competitors still cost at least $3/2$.
Thus it is $\gamma$-stable for every $1<\gamma<3/2$.
Nevertheless, single-linkage first merges the unique closest pair
$\{a,p\}$, crossing the optimal partition; subsequent merges cannot
repair this error. The metric perturbation increasing only $d(o,a)$ to $3/2$ makes
$\{\{o,b,c\},\{a,p\}\}$ tie the original optimum, so the
instance is not $\gamma$-stable for any $\gamma\ge3/2$.

Our algorithm builds on the partition-and-enumerate approach of
Gibson et al.~\cite{Gibson10}. Small clusters are recovered as
connected components. If few large clusters remain, we enumerate
them; otherwise we partition the instance, enumerate the few large
clusters split by the partition, and solve each residual region by
dynamic programming. The number guessed and the treewidth depend
only on $\eps$. One randomized run succeeds with probability at
least $1/3$; Section~\ref{sec:guarantee} proves Theorem~\ref{thm:main}
using the order family constructed in Appendix~\ref{app:derandomization}.

For unrestricted MSR, a quasi-polynomial-time approximation scheme
is known~\cite{Gibson10}, while a polynomial-time approximation scheme
(PTAS) remains open~\cite{GrandoniGY26}; the best known polynomial-time
guarantee is $3+\zeta$ for any fixed $\zeta>0$~\cite{BuchemERW24,GrandoniGY26}.
To our knowledge, the earlier examples of recovery under every fixed
stability factor $\gamma>1$ discussed in Section~\ref{sec:previous-work}
concern classes already admitting PTASs: clustering in fixed doubling
dimension~\cite{FriggstadKS25,FriggstadRS19}, metric
Max-Cut~\cite{BiluDLS13,FernandezKenyon01}, and planar independent
set~\cite{AngelidakisAABCD19,Baker94}.

We also prove coNP-completeness of recognizing stability and of
verifying stability of a supplied partition, for fixed rational $\gamma>1$.
Appendix~\ref{app:radius-costs} gives a corollary for power costs
$\sum_C r(C)^p$ with fixed $0<p\le1$.

After previous work, Section~\ref{sec:structure} proves the
structural results and Section~\ref{sec:algorithm} gives the recovery
algorithm. Sections~\ref{sec:recognition} and~\ref{sec:open-questions}
state the recognition result and open questions. Implementation
details and supporting proofs are in the appendices.

\section{Previous work}
\label{sec:previous-work}

Gibson et al.~\cite{Gibson10} proved MSR NP-hard even on planar
graph metrics and metrics of constant doubling dimension. Their
randomized exact algorithm takes $n^{O(\log n\log\Delta)}$ time,
where $\Delta$ is the aspect ratio, and yields a quasi-polynomial-time
approximation scheme. Its ball-cover formulation agrees with ours
by Lemma~\ref{lem:cover-conversion}.
In fixed-dimensional Euclidean metrics, they also give an exact
polynomial-time algorithm, assuming efficient comparisons of solution
costs~\cite{Gibson12}.
The general-metric algorithm randomly halves the diameter, enumerates
the $O(\log n)$ optimal balls cut with constant probability, and recurses for $O(\log\Delta)$
levels. Our density bounds and connectivity replace the partition
analysis's $O(\log n)$ harmonic factor by $O(\log(2/\eps))$.
At a suitable scale, only $O(\eps^{-1}\log(2/\eps))$ large clusters
need be guessed. The same density bounds give linear graph growth,
so bounded-treewidth dynamic programming replaces further recursive
partitioning. Both changes are needed to obtain an exponent
depending only on $\eps$.

The closest stability result is Gao and Reyzin's single-linkage
algorithm for $2$-stable MSR~\cite[Theorem~III.3]{GaoReyzin25}.
Below $2$, their separation bound weakens to $(\gamma-1)$ times
the cluster radius and no longer guarantees single-linkage recovery.
They also prove hardness below stability $2$ for Minimum
Sum-of-Diameters~\cite[Theorem~IV.2]{GaoReyzin25}.

For center-based clustering, recovery thresholds improved from
$3$~\cite{AwasthiBS12} to $1+\sqrt2$~\cite{BalcanLiang16} and
then $2$ under metric perturbations~\cite{AngelidakisMM17}.
For symmetric $k$-center, recovery below $2$ is hard unless
$\mathrm{NP}=\mathrm{RP}$, even under our arbitrary symmetric
perturbations~\cite[Theorem~10]{BalcanHW16}.
Ben-David and Reyzin~\cite{BenDavidReyzin14} establish hardness for
$k$-median and min-sum clustering under weaker structural stability
conditions.

Recovery for every fixed $\gamma>1$ is known for discrete $k$-means
and $k$-median in fixed-dimensional Euclidean and doubling
metrics~\cite{FriggstadKS25}. Those problems already admit
PTASs~\cite{FriggstadRS19}; their stability promise preserves the center
set, whereas ours preserves the partition.
Other examples include metric Max-Cut and a class of dense weighted
Max-Cut instances~\cite{BiluDLS13} and planar weighted independent
set~\cite{AngelidakisAABCD19}, also on classes admitting
PTASs~\cite{Baker94}.\footnote{Metric Max-Cut equivalently gives stable
recovery for two-cluster min-sum clustering: on each perturbed
instance, the within-cluster cost is the total pair weight minus the
cut weight. This two-cluster objective also has a
PTAS~\cite{FernandezKKR03}.}

Grandoni, Gupta, and Yadav~\cite{GrandoniGY26} give approximation
schemes for MSR with exponential dependence on $k$, leaving a PTAS
for variable $k$ open. Drexler et al.~\cite{DrexlerHKS23} give an XP
algorithm parameterized by treewidth for MSR on weighted graph metrics;
Kumar et al.~\cite{KumarMOS26} obtain fixed-parameter tractability
parameterized by treedepth for unweighted graph metrics. In those
results, graph distances define the input metric. Our graph instead
records interactions among candidate clusters, and stability bounds
its treewidth. Banerjee et al.~\cite{Banerjee25} study
parameterized radius objectives; Appendix~\ref{app:radius-costs}
uses their transformation $d\mapsto d^p$ to transfer stability.
Our remaining algorithmic ingredients are the linear-growth theorem
of Campbell et al.~\cite{Campbell23}, Korhonen's treewidth
approximation~\cite{Korhonen21}, and Indyk's approximately minwise
independent families~\cite{Indyk01}.

\section{Structural consequences of stability}
\label{sec:structure}

\paragraph{Computational model.}
Running times count exact arithmetic operations and comparisons,
with fixed parameters available as constants; random choices are
uniform integers. On inputs with rational distances of total binary
length $B_{\mathrm{in}}$, the same recovery guarantees hold for every
fixed real $\gamma>1$ with an additional
$\operatorname{poly}(B_{\mathrm{in}})$ factor in bit time: expected
for the randomized algorithm and worst-case for the deterministic
one. Appendix~\ref{app:full-correctness} gives the implementation.
The coNP statements use rational distances and fixed rational $\gamma$.

\paragraph{Notation.}
Throughout the paper, $1\le k\le n$ and $\eps=\gamma-1$.
In the recovery argument, triangle inequalities refer to the original
metric $d$. Balls $B(x,s)$ are closed, and $G_s$ joins distinct points
at distance at most $s$; $G_s[S]$ denotes its subgraph induced by
$S\subseteq P$. Write
$\dist(A,B)=\min_{x\in A,y\in B}d(x,y)$, with value $+\infty$ if
either set is empty, and $\dist(x,A)=\dist(\{x\},A)$.
We give the empty set diameter zero; radius statements concern
nonempty sets. We use fixed input order to break ties, and
$\log$ denotes the natural logarithm.

\subsection{Separation and the merge-and-split inequality}

Unless stated otherwise, $\OPT$ denotes the unique optimum of a
$\gamma$-stable instance. The results in this subsection hold for
every $\gamma>1$.

The next two lemmas hold for arbitrary metric instances, without
stability; their short proofs are in Appendix~\ref{app:elementary-proofs}.

\begin{lemma}
\label{lem:cover-conversion}
Every cover of $P$ by at most $k$ balls centered in $P$ can be converted
in polynomial time into an exact-$k$ partition of no greater cost.
\end{lemma}

\begin{lemma}
\label{lem:ball-budget}
For any optimal exact-$k$ partition $\OPT$ and every $x\in P$ and $s\ge0$,
\[
 \sum_{\substack{C\in\OPT\\C\subseteq B(x,s)}}r(C)\le s.
\]
\end{lemma}

The first inequality below is Gao and Reyzin's center-separation
bound. Appendix~\ref{app:elementary-proofs} gives a proof in our
exact-$k$, in-cluster-center model; its perturbation need not be metric.

\begin{lemma}[Gao--Reyzin {\cite[Lemma~A.1(1)]{GaoReyzin25}}]
\label{lem:separation}
Let $C\in\OPT$, let $c$ be any optimal center of $C$, and let
$p\notin C$. Then
\[
 d(c,p)>\gamma r(C).
\]
Consequently,
\[
 \dist(C,P\setminus C)>\eps r(C),\qquad
 C=B(c,r(C))=B(c,\gamma r(C)).
\]
\end{lemma}

The separation consequence follows from the triangle inequality:
for $x\in C$ and $p\notin C$,
$d(x,p)\ge d(c,p)-d(c,x)>\eps r(C)$.
The ball identities follow because all points of $C$ are within
$r(C)$ of $c$ and all other points are farther than $\gamma r(C)$.

\begin{lemma}
\label{lem:aligned}
Let $\OPT$ be a unique optimal exact-$k$ partition. If $W$ is a nonempty
union of its clusters, then $\OPT|_W$ is the unique optimum on $W$ for
its own number of clusters.
\end{lemma}

\begin{proof}
Extend any competing partition of $W$ by retaining the clusters of
$\OPT$ outside $W$. Their outside costs cancel, proving optimality
and uniqueness of the restriction.
\end{proof}

We next construct the perturbation used in the exchange argument.

\begin{lemma}[Split perturbation]
\label{lem:split}
For every nonsingleton set $C$ of radius $r$, there are an optimal
center $c\in C$, a point $p\in C\setminus\{c\}$, and a perturbation
that changes only distances within $C$ such that
\[
 r_{d'}(C)=\gamma r,
 \qquad r_{d'}(C\setminus\{p\})\le r.
\]
\end{lemma}

\begin{proof}
We seek an optimal center $c$ and a point $p$ such that every point
has a perturbed distance at least $\gamma r$ to another point,
while the distances from $c$ to $C\setminus\{p\}$ stay at most $r$.
The first condition forces every perturbed eccentricity to be at
least $\gamma r$; the second makes splitting off $p$ inexpensive.

Form a graph on $C$ by joining pairs at distance at least $r$.
Every vertex has a neighbor, since otherwise its eccentricity would
be below $r$. For an optimal center $c$ and a graph neighbor $p$,
suppose we keep $c$'s distances to $C\setminus\{p\}$ unchanged
and scale all other internal pairs by $\gamma$. The only possible
obstruction is a degree-one neighbor $v\ne p$ of $c$: its sole
distance of at least $r$ stays unscaled, making $v$ a center of
perturbed radius below $\gamma r$. We choose $c,p$ to avoid this.

Start with an optimal center $u$.
If $u$ has a degree-one neighbor $v$, choose $c=v$ and $p=u$.
Then $d(v,u)=r$ and all other distances from $v$ are below $r$,
so $c$ is optimal and its only graph edge is $\{c,p\}$.
Otherwise choose $c=u$ and any neighbor $p$ of $u$.

Leave $\{c,x\}$ unchanged for $x\ne p$, and multiply all other
internal distances by $\gamma$. In the first case, every graph
edge is multiplied by $\gamma$. In the second case, each vertex
other than $c,p$ has a graph edge not incident to $c$: a neighbor
of $c$ has degree at least two, and a nonneighbor has some edge.
The edge $\{c,p\}$ supplies the required distance for $c$ and $p$.
Thus every vertex has perturbed eccentricity at least $\gamma r$.
The optimal center $c$ attains this bound, while all its distances
to $C\setminus\{p\}$ remain unchanged and at most $r$.
\end{proof}

The constructed perturbation need not be metric: for
$x,y\in C\setminus\{c,p\}$, the scaled distance $\gamma d(x,y)$
may exceed the unchanged sum $d(x,c)+d(c,y)$. This is one obstacle
to extending the argument to metric perturbations
(Section~\ref{sec:open-questions}).

\begin{lemma}[Merge-and-split inequality]
\label{lem:merge}
Let $D_1,\ldots,D_s$, with $s\ge2$, be distinct optimal clusters.
Let $C_1,\ldots,C_{s-1}$ be distinct nonsingleton optimal clusters
outside this family. Then
\[
 r\left(\bigcup_{j=1}^sD_j\right)
 >\sum_{j=1}^s r(D_j)+\eps\sum_{i=1}^{s-1}r(C_i).
\]
\end{lemma}

\begin{proof}
Apply Lemma~\ref{lem:split} independently inside each $C_i$,
increasing the optimum's cost by $\eps\sum_i r(C_i)$. Merge the
$D_j$ and split the designated singleton from each $C_i$.
This removes and adds $s-1$ clusters, preserving the count $k$.
The merged set keeps its original radius; the remainder of each split cluster $C_i$
costs at most $r(C_i)$. The competitor therefore costs at most
\[
 \cost(\OPT)-\sum_jr(D_j)
       +r\left(\bigcup_jD_j\right).
\]
Uniqueness under the perturbation forces this upper bound to exceed
the optimum's perturbed cost, giving the inequality.
\end{proof}

\begin{corollary}
\label{cor:pairwise}
For distinct optimal clusters $C,D,E$, with $C$ nonsingleton,
\[
 \dist(D,E)+\min\{r(D),r(E)\}>\eps r(C).
\]
\end{corollary}

\begin{proof}
Assume $r(D)\ge r(E)$. Centering at an optimal center of $D$ and
traveling through a closest pair gives
\[
 r(D\cup E)\le r(D)+\dist(D,E)+2r(E).
\]
Combine this with Lemma~\ref{lem:merge}, splitting $C$.
\end{proof}

\subsection{Small clusters and local density}

For the rest of this section, assume $1<\gamma<2$ and fix a stable
instance with $k<n$.
Let
\[
 R=\max_{C\in\OPT}r(C),\quad b=\frac{\eps R}{4},\quad
 \mathcal T=\{C\in\OPT:r(C)\ge b\},\quad
 t=|\mathcal T|,\quad M=\sum_{C\in\mathcal T}r(C),\quad
 \rho=\frac{\eps M}{4}.
\]
Call the clusters in $\mathcal T$ \emph{large} and the others
\emph{small}. In particular, $t\ge1$ and $M\ge R>0$.
The factors $4$ leave room for both estimates used below:
$\eps R-b=3b>2b$ and $2\rho+2b\le\eps M$.
The radius $\rho$ is the range of the density estimate below;
regions of diameter at most $\rho$ inherit the resulting treewidth bound.

\begin{lemma}
\label{lem:small-components}
Every small optimal cluster is a connected component of the global
graph $G_{2b}$. Every other component consists entirely of points
belonging to large clusters.
\end{lemma}

\begin{proof}
A small cluster $D$ has diameter below $2b$, so it is connected in
$G_{2b}$. Choose an optimal cluster $C_*$ of radius $R$. For
$E\ne C_*,D$, Corollary~\ref{cor:pairwise} gives
\[
 \dist(D,E)>\eps R-r(D)>4b-b=3b.
\]
For $E=C_*$, separation gives
$\dist(D,C_*)>\eps R=4b$.
No edge leaves $D$, proving both claims.
\end{proof}

\begin{lemma}[Local cluster density]
\label{lem:density}
The following bounds hold for every $x\in P$.
\begin{enumerate}[label=(\roman*)]
\item A union of at least $t+1$ small optimal clusters has radius
      greater than $\eps M$.
\item For $0\le u\le\rho$, a ball $B(x,u)$ meets at most
      $1+(2u+2b)/(\eps b)$ small optimal clusters.
\item For every $u\ge0$, a ball $B(x,u)$ meets at most
      $(u+2R)/b$ large optimal clusters.
\end{enumerate}
\end{lemma}

\begin{proof}
For (i), choose an optimal center $z$ of the union and retain
$t+1$ clusters including the one containing $z$. The retained union
has no larger radius when centered at $z$. Lemma~\ref{lem:merge}, splitting all $t$ large
clusters, makes this radius greater than $\eps M$.

For (ii), let $W$ be the union of $s\ge1$ small clusters meeting
$B(x,u)$, and choose $z\in W\cap B(x,u)$. For $y\in W$, take
$y_0$ in its cluster's intersection with the ball. Then
\[
 d(z,y)\le d(z,x)+d(x,y_0)+d(y_0,y)<2u+2b.
\]
Thus $r(W)<2u+2b$, and
\[
 2u+2b\le2\rho+2b
   =\frac{\eps(M+R)}2\le\eps M.
\]
By (i), $s\le t$. For $s\ge2$, apply Lemma~\ref{lem:merge}
splitting $s-1$ large clusters, each of radius at least $b$:
\[
 \eps(s-1)b<r(W)<2u+2b.
\]
The asserted bound follows; $s=0,1$ is immediate.

For (iii), each meeting large cluster has diameter at most $2R$,
so it lies inside $B(x,u+2R)$. Lemma~\ref{lem:ball-budget} gives
\[
 b\,|\{C\in\mathcal T:C\cap B(x,u)\ne\varnothing\}|
 \le\sum_{\substack{C\in\mathcal T\\C\cap B(x,u)\ne\varnothing}}r(C)
 \le u+2R.
\]
This argument also covers the empty family.
\end{proof}

Lemma~\ref{lem:density} bounds the number of optimal clusters meeting
$B(x,u)$ by $O(\eps^{-2}(1+u/R))$ for $u\le\rho$.
Both the partition and treewidth arguments below use this linear growth.

\paragraph{Use of stability.}
Recovery below needs only a unique optimum, center separation
(Lemma~\ref{lem:separation}), and the merge-and-split inequality
(Lemma~\ref{lem:merge}). It therefore applies whenever these conditions
hold with $\gamma=1+\eps$ and $0<\eps<1$.

\section{A randomized algorithm for recovery}
\label{sec:algorithm}

For $\gamma\ge2$, run the algorithm below with parameter $3/2$:
every $\gamma$-stable instance is also $3/2$-stable.
Gao and Reyzin's single-linkage algorithm is a faster alternative
in this range~\cite[Theorem~III.3]{GaoReyzin25}. Henceforth $1<\gamma<2$ and
$\eps=\gamma-1$. We present the randomized algorithm first and
prove the deterministic theorem at the end of this section.

\paragraph{Proof of Theorem~\ref{thm:main} at a glance.}
The argument has three main dependencies:
\begin{itemize}
\item \textbf{Small-cluster recovery.} Merge-and-split
(Lemma~\ref{lem:merge}) gives separation
(Corollary~\ref{cor:pairwise}), which identifies small clusters as
components (Lemma~\ref{lem:small-components}).
\item \textbf{Few split large clusters.} Density
(Lemma~\ref{lem:density}) bounds net sizes and splitting probabilities
(Lemmas~\ref{lem:net-counts} and~\ref{lem:cut} in
Appendix~\ref{sec:partition-construction}), yielding
Proposition~\ref{prop:partition}.
\item \textbf{Local recovery.} Density and the linear-growth
theorem (Theorem~\ref{thm:linear-growth}) bound treewidth
(Lemma~\ref{lem:treewidth}), enabling the local solver
(Proposition~\ref{prop:local}).
\end{itemize}
Theorem~\ref{thm:randomized-recovery} combines these steps.
The order family of Theorem~\ref{thm:minwise-family} then removes
randomness, as shown at the end of Section~\ref{sec:guarantee}.

\subsection{The search and its two branches}

Either there are few large clusters, so we enumerate them, or their
total radius is large enough to support a useful partition scale.
Fix $\sigma=2^9$, a power of two large enough for the net-count
lemma (Lemma~\ref{lem:net-counts}), as explained in
Appendix~\ref{app:search-details}.

\begin{table}[H]
\centering
\small
\caption{The four scales used by the recovery algorithm.}
\label{tab:scales}
\begin{tabularx}{\linewidth}{@{}lX@{}}
\toprule
Scale & Role \\
\midrule
$R$ & Largest optimal radius; guessed from input distances. \\
$b=\eps R/4$ & Threshold between small and large clusters. \\
$\rho=\eps M/4$ & Region diameter bound for density and treewidth; $M$ is the total large-cluster radius. \\
$L$ & Partition scale, chosen so $\sigma R\le L\le\rho<2L$. \\
\bottomrule
\end{tabularx}
\end{table}

\begin{itemize}
\item \textbf{Few large clusters ($\rho<\sigma R$).}
There are at most $M/b=O(\eps^{-2})$ large clusters. Enumerate them
and recover the remaining small clusters as connected components.
\item \textbf{Partition branch ($\rho\ge\sigma R$).}
Choose $L$ with $\sigma R\le L\le\rho<2L$. With constant probability,
the partition splits only $O(\eps^{-1}\log(2/\eps))$ large clusters.
Enumerate those clusters and solve the residual regions.
\end{itemize}

In the partition branch, $\rho$ is where partitioning gives way to
dynamic programming. The small-cluster density bound in
Lemma~\ref{lem:density}(ii) is established only for radii at most
$\rho$, so our treewidth argument applies to aligned regions of
diameter at most $\rho$. Recursing from $\rho$ down to $R$ could
require $\Theta(\log(\rho/R))$ levels; since $\rho/R$ can grow with
$n$, repeating the enumeration at each level could lead to a
quasi-polynomial running time. Instead, one partition at
$L\approx\rho$ has $M/L=O(1/\eps)$: it splits few large clusters
and leaves residual regions small enough for the treewidth argument.

The component test in Lemma~\ref{lem:small-components} is one-sided:
every small optimal cluster is a global $G_{2b}$-component of radius
below $b$, but a large cluster need not be connected in $G_{2b}$.
Thus a component passing the test may instead contain pieces of
one or more large clusters. The partition keeps these components whole, while
the local solver decides whether to retain each as a cluster or
cover it completely with large-cluster candidates.

The algorithm guesses $R$ among the positive input distances and
runs both branches. For a guess $\widehat R$, put
$\widehat b=\eps\widehat R/4$. A candidate is a ball $A=B(c,s)$
with a specified center $c$ and radius $s$ satisfying
\begin{equation}
\label{eq:candidates}
 s\in\{d(c,x):x\in P\},\quad
 \widehat b\le s\le\widehat R,\quad
 \widehat b\le r(A)\le\widehat R,\quad
 B(c,s)=B(c,\gamma s).
\end{equation}
All tests use the full metric, and the candidate's cost is $r(A)$.
There are at most $n^2$ representations. At $\widehat R=R$, every
large optimal cluster occurs, by Lemma~\ref{lem:separation}.
Only disjoint candidate families are selected, so multiple
representations of one set cannot be selected together.

Write $N_\eps=O(\eps^{-2})$ and
$F_\eps=O(\eps^{-1}\log(2/\eps))$ for the two enumeration bounds;
their exact values appear in Table~\ref{tab:parameters}.
The routine $\texttt{RandomPartitions}(\widehat R,L)$ returns
$O(n^2)$ partitions into regions of diameter at most $L$.
The routine $\texttt{SolveRegion}(\widehat R,W)$ either rejects
or returns a table $\DP_W(j)$ of feasible $j$-cluster partitions of
$W$ and their actual costs. Unavailable entries are $+\infty$;
$\DP_\varnothing(0)=0$.

\begin{algorithm}[H]
\caption{Randomized recovery of a stable MSR clustering}
\label{alg:recovery}
\small
\begin{tabularx}{\linewidth}{@{}r@{\quad}>{\raggedright\arraybackslash}X@{}}
1 & If $k=n$, return the singleton partition; otherwise initialize any feasible exact-$k$ answer.\\
2 & For every positive input distance $\widehat R$, form the candidates in~\eqref{eq:candidates} and run both branches:\\
3 & \textbf{A (few large clusters).} Enumerate disjoint candidate families $\mathcal X$ with $|\mathcal X|\le\min\{N_\eps,k\}$. Combine $\mathcal X$ with the components of $G_{2\widehat b}[P\setminus\bigcup\mathcal X]$.\\
4 & \textbf{B (partition branch).} For every $L=2^j\widehat R$, $\log_2\sigma\le j\le\lceil\log_2n\rceil$, and every $\mathcal Q\in\texttt{RandomPartitions}(\widehat R,L)$, enumerate disjoint candidate families $\mathcal X$ with $|\mathcal X|\le\min\{F_\eps,k\}$.\\
5 & For each such choice, set $W_Q=Q\setminus\bigcup\mathcal X$ and call $\texttt{SolveRegion}(\widehat R,W_Q)$ for all $Q\in\mathcal Q$. If none rejects, combine their tables using~\eqref{eq:combine-regions}.\\
6 & Retain only feasible exact-$k$ proposals, updating the answer whenever its actual cost decreases. Return the cheapest answer.\\
\end{tabularx}
\end{algorithm}

The table combination minimizes
\begin{equation}
\label{eq:combine-regions}
 \sum_{A\in\mathcal X}r(A)+\sum_{Q\in\mathcal Q}\DP_{W_Q}(j_Q)
 \quad\text{subject to}\quad
 \sum_{Q\in\mathcal Q}j_Q=k-|\mathcal X|.
\end{equation}
The table combination takes polynomial time by adding tables one at a time.
Each partition is reused for all candidate families at its scale;
each call to $\texttt{RandomPartitions}$ uses a fresh ordering.
The detailed implementations and feasibility checks are in
Appendix~\ref{app:recovery-details}.

\subsection{Partitions that split few clusters}
\label{sec:partition-overview}

A set is \emph{split} by a partition if it meets more than one part.

\begin{proposition}
\label{prop:partition}
For every $\widehat R>0$ and $L\ge\sigma\widehat R$,
$\texttt{RandomPartitions}(\widehat R,L)$ returns in polynomial
time a nonempty family of $O(n^2)$ partitions into nonempty regions
of diameter at most $L$. On a stable input at $\widehat R=R$, every
partition preserves all small optimal clusters. If also
$\sigma R\le L\le\rho<2L$, then with probability at least $1/3$
the family contains a partition with at most $F_\eps$ split large
clusters.
\end{proposition}

Here is why the splitting bound depends on $\eps$ rather than $n$.
Keep components of $G_R$ of diameter at most $L$ whole. In every
remaining component $U$, choose a maximal strictly $2R$-separated
set $N_U$. Process these points in random order, assigning to each
the unassigned points within a common radius
$\beta\in[L/16,L/8]$. A final adjustment keeps every global
$G_{2b}$-component of radius below $b$ whole. The implementation
enumerates $O(n^2)$ midpoint radii, so only the ordering is random.

For the analysis, choose $\beta$ uniformly from its interval.
Fix a large cluster $C$ of radius $r$ in $U$ and an optimal center $c$.
Let $S_C$ be $C$ together with every global $G_{2b}$-component of
radius below $b$ that meets $C$. Keeping $S_C$ whole before the
adjustment ensures that $C$ remains whole afterward. Moreover,
$S_C\subseteq B(c,3r)$ and $\diam(S_C)\le6r$.

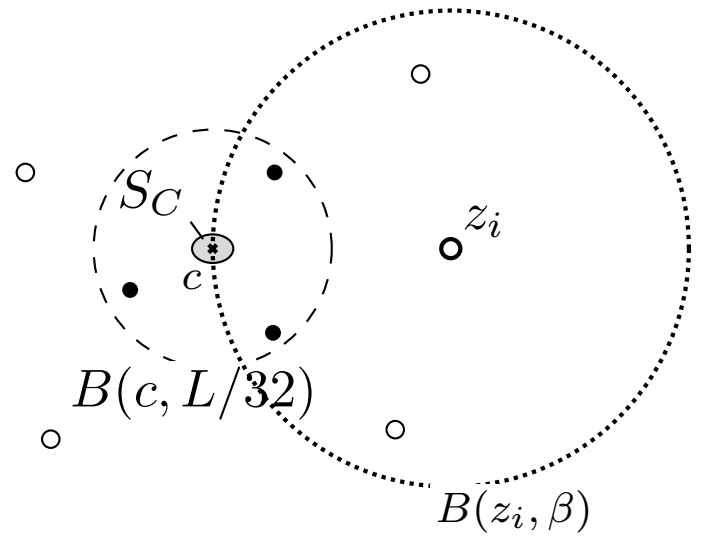
\begin{figure}[htbp]
\centering
\begin{minipage}[c]{0.41\linewidth}
\centering
\begin{tikzpicture}[x=1.05cm,y=1.05cm,font=\small]
  \path[use as bounding box] (-1.55,-1.85) rectangle (3.15,1.85);
  \draw[dashed] (0,0) circle (0.75);
  \node[fill=white,inner sep=1pt] at (-0.12,-0.91) {$B(c,L/32)$};
  \filldraw[fill=gray!30,draw=black] (0,0) ellipse (0.13 and 0.09);
  \node[anchor=south east,inner sep=1pt] at (-0.13,0.17) {$S_C$};
  \draw[thin] (-0.14,0.17) -- (-0.06,0.06);
  \draw[thick] (-0.025,-0.025) -- (0.025,0.025)
               (-0.025,0.025) -- (0.025,-0.025);
  \node[font=\scriptsize,anchor=north east,inner sep=1pt] at (-0.03,-0.10) {$c$};
  \fill (-0.52,-0.26) circle (0.05);
  \fill (0.39,0.48) circle (0.05);
  \fill (0.38,-0.53) circle (0.05);
  \draw[fill=white] (-1.18,0.48) circle (0.055);
  \draw[fill=white] (-1.02,-1.20) circle (0.055);
  \draw[fill=white] (1.15,-1.14) circle (0.055);
  \draw[fill=white] (1.31,1.10) circle (0.055);
  \draw[densely dotted,thick] (1.50,0) circle (1.50);
  \draw[fill=white,thick] (1.50,0) circle (0.06);
  \node[anchor=south west,inner sep=2pt] at (1.50,0) {$z_i$};
  \node[fill=white,inner sep=1pt,anchor=north,font=\scriptsize]
    at (1.90,-1.48) {$B(z_i,\beta)$};
\end{tikzpicture}
\end{minipage}\hfill
\begin{minipage}[c]{0.57\linewidth}
\small
\textbf{Inner centers} (filled dots) lie in $B(c,L/32)$.
Their radius-$\beta$ balls always cover $S_C$, since
$\beta\ge L/32+3r$.

\medskip
\textbf{Relevant outer centers} (open dots) lie outside that ball
and within distance $L/8$ of $S_C$. They lie in $B(c,L/4)$
(not drawn), and their balls may meet only part of $S_C$.
Only a partial intersection at the first ball meeting $S_C$ can
split it.
\end{minipage}
\caption{The partition analysis inside a wide component $U$.
Filled and open dots represent points of $N_U$; $S_C\subseteq B(c,3r)$.
The first center meeting $S_C$ splits it only if its radius-$\beta$
ball fails to cover it, as illustrated by $z_i$.
All radii are drawn to scale with $R=L/512$ and $\beta=L/16$.}
\label{fig:partition-analysis}
\end{figure}

Figure~\ref{fig:partition-analysis} distinguishes two kinds of net
centers. Connectivity supplies $\Omega(L/R)$ \emph{inner centers}
in $B(c,L/32)$, whose balls cover $S_C$ for every allowed radius.
The density bound limits all relevant centers to
$O(L/(\eps^2R))$ (Lemma~\ref{lem:net-counts}). The ratio of outer
to inner centers is therefore $O(\eps^{-2})$, independent of $n$.

If $S_C$ is split, the first center whose ball meets it must meet
it only partially. Averaging over the radius contributes a factor
$O(r/L)$; summing over the possible first centers gives a harmonic
factor controlled by the outer-to-inner ratio. Consequently,
\[
 \Pr[C\text{ is split}]=O\bigl((r/L)\log(2/\eps)\bigr).
\]
Since $M/L=O(1/\eps)$, summing this bound and applying Markov's
inequality gives the claimed success probability.
Appendix~\ref{sec:partition-construction} proves the bound and explains
why enumerating midpoint radii preserves it.

\subsection{Density, treewidth, and local recovery}
\label{sec:local-overview}

The local solver uses two properties: \textbf{alignment}, meaning
that $W$ is a union of whole optimal clusters, and \textbf{bounded
diameter}, meaning $\diam(W)\le\rho$. Together they guarantee
\textbf{complete local recovery}.

\begin{proposition}
\label{prop:local}
For every input, $\texttt{SolveRegion}(\widehat R,W)$ halts in
$n^{O(\eps^{-5})}$ time and either rejects or returns feasible
partitions with their actual costs. If the input is stable,
$\widehat R=R$, and $W$ is aligned with $\diam(W)\le\rho$, it does
not reject, and its table contains $\OPT|_W$ at the true cluster count.
\end{proposition}

Choose a maximal strictly $2\widehat R$-separated set $N_W\subseteq W$
and join two vertices when their distance is at most $8\widehat R$;
call the resulting graph $\mathcal H_W$. Assign each candidate
contained in $W$ to the vertex nearest its specified center.
At the correct guess, each true large cluster assigned to $v$ lies
in $B(v,3R)$. Lemma~\ref{lem:ball-budget} bounds their total radius
by $3R$, so at most $3R/b=12/\eps$ such clusters are assigned to
any vertex. Selecting one representation per cluster, we can
therefore restrict each variable to $O(1/\eps)$ selected candidates,
giving $n^{O(1/\eps)}$ possible values.

The constraints enforce disjointness and coverage. A global
$G_{2\widehat b}$-component contained in $W$ with radius below
$\widehat b$ is \emph{optional}. Because the component test is
one-sided, the dynamic program either retains it as a cluster or
covers it completely using selected large-cluster candidates.
All other points must be covered. Every constraint involves a clique
of $\mathcal H_W$, and each variable
contributes its actual cost and cluster count. These facts, including
their validity on arbitrary inputs, are proved in
Appendix~\ref{app:recovery-details}.

\begin{theorem}[Campbell et al. {\cite[Theorem~1]{Campbell23}}]
\label{thm:linear-growth}
If $\kappa>1$ and, for every integer $j\ge1$, every connected subgraph
of a graph $H$ having graph radius at most $j$, measured within that
subgraph, has at most $\kappa j$ vertices, then
$\tw(H)\le49\kappa^2+30\kappa$.
\end{theorem}

\begin{lemma}
\label{lem:treewidth}
On a stable input at $\widehat R=R$, an aligned region $W$ with $\diam(W)\le\rho$
satisfies $\tw(\mathcal H_W)\le\omega_\eps=O(\eps^{-4})$.
\end{lemma}

\begin{proof}
Distinct vertices of $N_W$ belong to distinct optimal clusters,
whose diameters are at most $2R$. For $j\ge1$, a ball of graph radius $j$ lies in
the metric ball of radius $u=\min\{8jR,\diam(W)\}\le\rho$.
Using $u\le8jR$ and $b=\eps R/4$, Lemma~\ref{lem:density} bounds its size by
\[
 \frac{u+2R}{b}+1+\frac{2u+2b}{\eps b}
 \le1+\frac{32j+10}{\eps}+\frac{64j}{\eps^2}
 \le c_\eps j,
\]
where $c_\eps=1+80/\eps+80/\eps^2$.
Every connected subgraph of radius at most $j$ lies in a ball of
radius $j$ in $\mathcal H_W$. Theorem~\ref{thm:linear-growth} gives the
bound with $\omega_\eps=\lceil49c_\eps^2+30c_\eps\rceil$.
\end{proof}

A net restricted to points in large optimal clusters, if it preserved
the solver's coverage guarantees, would have growth constant
$O(\eps^{-1})$ and treewidth $O(\eps^{-2})$, reducing the running time
to $n^{O(\eps^{-3})}$. The available tests do not identify the
required points: the component test can accept pieces of large clusters, while the
candidate tests can accept unions of small ones. The current
construction therefore spans all of $W$ and uses both density bounds.

\paragraph{Proof sketch of Proposition~\ref{prop:local}.}
Korhonen's algorithm~\cite[Theorem~I.1]{Korhonen21} supplies a
decomposition of width at most $2\omega_\eps+1$, or correctly rejects
because treewidth exceeds $\omega_\eps$. Recording candidate choices
and cluster counts in each bag gives $n^{O(\eps^{-5})}$ time.
Alignment and the candidate restriction preserve the local optimum.
Appendix~\ref{sec:local-construction} gives the full proof;
Appendix~\ref{app:dp} gives the recurrence.

\subsection{Correctness and running time}
\label{sec:guarantee}

\begin{theorem}[Randomized recovery]
\label{thm:randomized-recovery}
Algorithm~\ref{alg:recovery} always returns a feasible exact-$k$
partition in $n^{O(\eps^{-5})}$ time.
On a $\gamma$-stable input it returns the optimum
with probability at least $1/3$. Taking the cheapest result of $m$
independent runs reduces the failure probability to $(2/3)^m$.
\end{theorem}

\paragraph{Proof sketch of Theorem~\ref{thm:randomized-recovery}.}
Every accepted proposal is feasible. On a stable input at the correct
guess, Branch A enumerates all large clusters when $\rho<\sigma R$;
Lemma~\ref{lem:small-components} recovers the rest. Otherwise
Lemma~\ref{lem:scale-dichotomy} supplies a scale
$\sigma R\le L\le\rho<2L$. With probability at least $1/3$,
Proposition~\ref{prop:partition} splits at most $F_\eps$ large clusters.
Enumerating and removing them leaves aligned regions of diameter at
most $L\le\rho$. Proposition~\ref{prop:local} and
\eqref{eq:combine-regions} then recover $\OPT$.
Only this one partition call needs to succeed. Local computation
dominates the running time. Appendix~\ref{app:full-correctness}
gives the full proof.

\begin{proof}[Proof of Theorem~\ref{thm:main}]
If $k=n$, return the singleton partition.
It suffices to treat $1<\gamma<2$: for $\gamma\ge2$, use parameter
$3/2$ and monotonicity of stability.
Theorem~\ref{thm:minwise-family}, proved from Indyk's theorem in
Appendix~\ref{app:derandomization}, supplies a polynomial-size family
$\mathcal F_V$ of orders of the net points $V$ used by a partition call.
Under a uniform choice from this family, each element of a nonempty
$T\subseteq V$ is first in $T$ with probability at most $3/(2|T|)$.
Replace the sampled order in each partition call by enumeration of
$\mathcal F_V$, keeping the same midpoint radii and all other steps.
When $V=\varnothing$, keep the single partition already specified.

On a stable input with $\rho<\sigma R$, Branch A already succeeds
deterministically. Otherwise take the correct guess and scale from
the proof sketch above. If $V=\varnothing$, no cluster is split.
For nonempty $V$, apply the proof of Lemma~\ref{lem:cut} with
$1/(n_{\mathrm{in}}+i)$ replaced by $3/(2(n_{\mathrm{in}}+i))$.
Averaging over the order family and the independent auxiliary radius
then gives $\mathbb E Z<F_\eps$ by the factor-$3/2$ slack built into
$F_\eps$ (Table~\ref{tab:parameters}).
Some enumerated order and midpoint radius therefore split at most
$F_\eps$ large clusters, and Theorem~\ref{thm:randomized-recovery}'s
recovery argument applies to this partition.

Enumerating $n^{O(1)}$ orders adds only a polynomial factor.
Feasibility and the running-time bounds hold for every order and
every input, so the resulting deterministic algorithm always halts
with a feasible exact-$k$ partition in $n^{O(\eps^{-5})}$ time.
\end{proof}

Returning the cheaper of this partition and a deterministic $(3+\zeta)$-approximate
solution~\cite{BuchemERW24}, converted using Lemma~\ref{lem:cover-conversion},
preserves exact recovery on stable inputs and gives a $(3+\zeta)$
guarantee on every input, for any fixed $\zeta>0$.

\section{Recognizing stability}
\label{sec:recognition}

\emph{Recognition} asks whether an instance is $\gamma$-stable;
\emph{specified-partition verification} asks whether a given partition
remains uniquely optimal under every allowed perturbation. We also
consider \emph{metric perturbations}, which must satisfy the triangle
inequality. Centers may change.

\begin{samepage}
\begin{theorem}
\label{thm:recognition-hardness}
For every fixed rational $\gamma>1$, on rational finite metric inputs:
\begin{enumerate}[label=(\roman*)]
\item Specified-partition verification is coNP-complete, both for
arbitrary symmetric perturbations and for metric perturbations.
\item Recognition is coNP-complete for arbitrary symmetric perturbations
and coNP-hard for metric perturbations.
\end{enumerate}
In all cases, the complement of exact MSR threshold decision reduces
to the stability problem by a polynomial-time many-one reduction.
The reduction supplies the unique unperturbed optimum, which has
exactly one nonsingleton cluster.
\end{theorem}
\end{samepage}

Recovery assumes stability; recognition tests it. The reduction supplies
the unique unperturbed optimum: the source instance and a hub form one
cluster, with dummy points as singletons. A competitor merges the dummies,
isolates the hub, and partitions the source instance. Increasing the
hub-to-source distances creates a tie exactly when the source has a
sufficiently cheap solution.

Deterministic recovery removes recognition's outer existential choice
of an optimal partition: it produces a candidate in polynomial time
on every input, equal to the stable optimum whenever one exists.
The input is therefore unstable exactly when this candidate has an
instability certificate, polynomially checkable by
Lemma~\ref{lem:instability-certificate}. This gives coNP membership
for arbitrary symmetric perturbations. The same argument is unavailable
for metric perturbations, where recovery remains open.
Appendix~\ref{app:recognition-proof} gives the proofs.

\section{Open questions}
\label{sec:open-questions}

Can the smaller net discussed in Section~\ref{sec:local-overview}
reduce the running time to $n^{O(\eps^{-3})}$? More broadly, can the
running time be made $g(1/\eps)\,n^{O(1)}$, or can this be ruled out,
for example by W[1]-hardness parameterized by $\lceil1/\eps\rceil$
under reductions that preserve the stability promise?
Can the extra $\eps^{-1}$ factor in the exponent from storing candidate
families be avoided? Letting each vertex of a finer net record at most one
candidate covering it would avoid families, but a large cluster may
contain arbitrarily many such vertices. Their choices must agree,
and they form a clique under our current interaction rule, giving
unbounded treewidth.
Recovery under stability against metric perturbations alone remains
open. Finally, the different thresholds for MSR and MSD suggest studying
which cluster costs allow a split that removes the cluster's
perturbation increase while preserving its center.

\section*{Acknowledgments}

This work was supported in part by NSF grant ECCS-2217023.

\section*{AI methodology}

We used \texttt{Claude Fable 5.1}, \texttt{Claude Opus 5.5}, \texttt{ChatGPT 6}, and \texttt{Gemini 4 Argon}
to assist with developing and refining the mathematical results and
with writing the paper. The authors selected the problem, identified
the unresolved regime $1<\gamma<2$, directed parts of the research
strategy, and decided which results and arguments to retain.
\texttt{Claude Opus 5.5} proposed the central proof idea
after identifying failures in lower-bound approaches that the authors
had asked it to investigate. All original theorems and lemmas were
either proposed by these tools or refined with their assistance.
The tools materially influenced all sections, including the appendices.
The authors primarily determined the presentation and exposition,
with AI assistance in drafting and revising the text. The authors
verified the correctness and originality of all content, including
references, and take full responsibility for the content of the paper.

\appendix
\section{Supporting details for recovery}
\label{app:recovery-details}

\subsection{Elementary structural facts}
\label{app:elementary-proofs}

\begin{proof}[Proof of Lemma~\ref{lem:cover-conversion}]
Keep only the largest ball at each repeated center. Assign each
remaining center to its own cluster and every other point to a ball
containing it. Each cluster has radius at most its ball's radius.
While there are fewer than $k$ clusters, choose a nonsingleton cluster,
retain an optimal center of that current cluster, and split off another
point as a singleton. Such a cluster exists because $k\le n$, and
retaining its center prevents any cost increase.
\end{proof}

\begin{proof}[Proof of Lemma~\ref{lem:ball-budget}]
For a nonempty family $\mathcal F$ in the sum, replace its
center-radius balls by $B(x,s)$ and retain those of the other clusters.
The resulting $k-|\mathcal F|+1\le k$ balls cover $P$, regardless of
any overlaps, at total cost
\[
 \cost(\OPT)-\sum_{C\in\mathcal F}r(C)+s.
\]
Convert this cover by Lemma~\ref{lem:cover-conversion} and apply
optimality. The empty family satisfies the bound trivially.
\end{proof}

\begin{proof}[Proof of Lemma~\ref{lem:separation}]
Suppose $d(c,p)\le\gamma r(C)$ for $p\in C'\ne C$.
Then $r(C)>0$. Multiply every distance by $\gamma$, except the
pairs $\{c,y\}$ with $y\in C'$, which remain unchanged.
Every intra-cluster distance of $\OPT$ is scaled, so
$\cost_{d'}(\OPT)=\gamma\cost_d(\OPT)$.
We construct a different exact-$k$ partition of no greater cost.

If $C'=\{p\}$, choose $x\in C\setminus\{c\}$ and replace
$C,C'$ by $(C\setminus\{x\})\cup\{p\}$ and $\{x\}$.
Centered at $c$, the nonsingleton set has radius at most
$\max\{\gamma r(C),d(c,p)\}=\gamma r(C)$.

If $C'$ has an optimal center other than $p$, replace $C,C'$ by
$C\cup\{p\}$ and $C'\setminus\{p\}$. Keeping $c$ and that
other center gives radii at most $\gamma r(C)$ and $\gamma r(C')$.

In the remaining case, $p$ is the unique optimal center of $C'$ and $|C'|\ge2$.
Replace $C,C'$ by $C\cup(C'\setminus\{p\})$ and $\{p\}$.
For $y\in C'\setminus\{p\}$,
\[
 d'(c,y)=d(c,y)\le d(c,p)+d(p,y)
 \le\gamma r(C)+r(C').
\]
Thus the replacement costs at most $\gamma r(C)+r(C')$, saving
at least $(\gamma-1)r(C')>0$ compared with the two original clusters
under $d'$. All three cases contradict uniqueness under $d'$.
\end{proof}

\subsection{Parameters and enumeration bounds}
\label{app:search-details}

The exact bounds used by the algorithm are listed below. Fix
$\sigma=512=2^9$. This choice is a power of two, so the doubling
list of partition scales contains a value in $[\sigma R,\rho]$
whenever $\rho\ge\sigma R$, as shown in
Lemma~\ref{lem:scale-dichotomy}; the scale index therefore starts
at $9$. The lower bound in Lemma~\ref{lem:net-counts} uses
$\sigma\ge480$. Set $\chi=16\cdot6=96$: the factors come from
the radius interval of length $L/16$ and the bound
$\diam(S_C)\le6r$ in Lemma~\ref{lem:cut}. Use the abbreviations
\[
 \Lambda_\eps=\left\lceil\log_2\left(1+\frac{12\sigma}{\eps^2}\right)\right\rceil,
 \qquad c_\eps=1+\frac{80}{\eps}+\frac{80}{\eps^2}.
\]
\begin{table}[H]
\centering
\small
\caption{Bounds used in the search and local computation.}
\label{tab:parameters}
\begin{tabularx}{\linewidth}{@{}llX@{}}
\toprule
Parameter & Value & Role \\
\midrule
$N_\eps$ & $\lceil16\sigma/\eps^2\rceil$ & Clusters enumerated in Branch A \\
$F_\eps$ & $\lceil12\chi\Lambda_\eps/\eps\rceil$ & Clusters enumerated in Branch B \\
$h_\eps$ & $\lfloor12/\eps\rfloor$ & Selected candidates at each vertex \\
$\omega_\eps$ & $\lceil49c_\eps^2+30c_\eps\rceil$ & Treewidth threshold \\
\bottomrule
\end{tabularx}
\end{table}

\subsection{Partition construction}
\label{sec:partition-construction}

The routine $\texttt{RandomPartitions}(\widehat R,L)$ takes
$L\ge\sigma\widehat R$. Keep every component of $G_{\widehat R}$
of diameter at most $L$ as one region. Call each remaining component
\emph{wide}. In each wide component $U$, greedily choose a maximal
strictly $2\widehat R$-separated set
$N_U\subseteq U$, using input order. Strict separation means that
distinct points of $N_U$ have distance greater than $2\widehat R$. Put
\[
 V=\bigcup_{\substack{U\text{ a component of }G_{\widehat R}\\
                      \diam(U)>L}}N_U.
\]
If $V=\varnothing$, return the single partition consisting of the
components of $G_{\widehat R}$.

The final adjustment ensures that small clusters move whole:
at the correct guess, they are global $G_{2b}$-components of radius
below $b$. For each global $G_{2\widehat b}$-component $K$ with
$r(K)<\widehat b$, fix a representative $a_K\in K$ by input order.
For an order $\pi$ of $V$ and $\beta\in[L/16,L/8]$, define
$\texttt{Partition}(\widehat R,L;\pi,\beta)$ as follows.
In each wide component $U$, process $N_U$ in the restriction of
$\pi$. Assign to each center the unassigned points of $U$ within
distance $\beta$, producing the \emph{initial parts}. Give the components kept
whole their own initial labels. Then simultaneously give every point
of each component $K$ with $r(K)<\widehat b$ the initial label of
its representative $a_K$. Discard empty labels. The sets $N_U$ and
representatives $a_K$ are fixed before choosing $\pi$ or $\beta$.

Form the breakpoints
\[
 \mathcal B=\{L/16,L/8\}\ \cup\
 \{d(z,x):z\in V,\ x\in P,\ L/16<d(z,x)<L/8\}.
\]
Sort the distinct values as $\beta_0<\cdots<\beta_a$ and let
\[
 \mathcal R_L=\{(\beta_i+\beta_{i+1})/2:0\le i<a\}.
\]
There are at most $n^2+1$ radii. Choose a uniformly random permutation
$\pi$ of $V$ and enumerate
\[
 \texttt{Partition}(\widehat R,L;\pi,\beta),
 \qquad \beta\in\mathcal R_L,
\]
using the same sampled order $\pi$ for every radius. Repeated
partitions may be retained. All choices other than $\pi$ are deterministic.

The next lemma combines a lower bound from connectivity with an
upper bound from local density. Their ratio depends only on $\eps$;
this is what makes the harmonic factor in the subsequent splitting
bound independent of $n$.

\begin{lemma}
\label{lem:net-counts}
On a stable input with $\widehat R=R$ and
$\sigma R\le L\le\rho$, for every wide component $U$ and $x\in U$,
\[
 |N_U\cap B(x,L/32)|\ge\frac{L}{\sigma R},
 \qquad
 |N_U\cap B(x,L/4)|\le\frac{12L}{\eps^2R}.
\]
\end{lemma}

\begin{proof}
Since $\diam(U)>L$, some $y\in U$ satisfies $d(x,y)>L/2$.
Follow a $G_R$-path from $x$ to $y$; distance from $x$ changes by
at most $R$ per edge. For
\[
 j=1,\ldots,\left\lfloor\frac{L}{(\sigma/2)R}\right\rfloor,
\]
we have $6Rj\le12L/\sigma<L/2$. Let $v_j$ be the first path
vertex at distance at least $6Rj$ from $x$. Its distance is at most
$6Rj+R$. Maximality supplies a point $z_j\in N_U$ within $2R$ of $v_j$,
so
\[
 6Rj-2R\le d(x,z_j)\le6Rj+3R.
\]
These intervals are disjoint, so the $z_j$ are distinct. Their upper
endpoints satisfy $12L/\sigma+3R\le15L/\sigma\le L/32$ because
$L\ge\sigma R$ and $\sigma=512\ge480$.
Their number is $\lfloor L/((\sigma/2)R)\rfloor\ge L/(\sigma R)$.

For the upper bound, different points of $N_U$ belong to different
optimal clusters. Applying Lemma~\ref{lem:density} at $u=L/4\le\rho$
gives
\begin{align*}
 |N_U\cap B(x,L/4)|
 &\le\frac{L/4+2R}{b}+1+\frac{L/2+2b}{\eps b}\\
 &=\frac{L}{\eps R}+\frac{2L}{\eps^2R}
      +\frac{10}{\eps}+1
 \le\frac{12L}{\eps^2R}.
\end{align*}
The last inequality uses $0<\eps<1$ and $L\ge\sigma R=512R$.
\end{proof}

For a large cluster $C$ of radius $r$, define
\[
 S_C=C\ \cup\!
 \bigcup_{\substack{K\text{ a component of }G_{2b}\\
                    r(K)<b,\ K\cap C\ne\varnothing}}K.
\]
If $c$ is an optimal center of $C$, then
\[
 S_C\subseteq B(c,r+2b)\subseteq B(c,3r),
 \qquad \diam(S_C)\le6r,
\]
since every added point is within $2b$ of $C$ and $r\ge b$.
If $S_C$ lies in one initial part, every representative affecting $C$
has that label, so the adjustment leaves $C$ whole.

For the analysis, let $\pi$ be the uniform random permutation chosen
by the routine and independently choose $\beta$ uniformly from
$[L/16,L/8]$. This auxiliary random radius is used only in the proof;
the algorithm enumerates the midpoint radii in $\mathcal R_L$.

\begin{lemma}
\label{lem:cut}
Suppose $V\ne\varnothing$, $1<\gamma<2$, and the input is
$\gamma$-stable. If
$\widehat R=R$ and $\sigma R\le L\le\rho$, then every large cluster
$C$ of radius $r$ satisfies
\[
 \Pr[C\text{ is split by the final partition}]
 \le\frac{\chi r}{L}\log\left(1+\frac{12\sigma}{\eps^2}\right).
\]
Hence, if $Z$ is the number of split large clusters,
\[
 \mathbb E Z\le\frac{\chi M}{L}
       \log\left(1+\frac{12\sigma}{\eps^2}\right).
\]
\end{lemma}

\begin{proof}
A cluster in a component kept whole cannot split. Otherwise, let $U$
be its component, $S=S_C$, $\delta_C=\diam(S)\le6r$, and $c$ an
optimal center of $C$. Every center in $N_U\cap B(c,L/32)$ covers $S$
for every allowed $\beta$, since
\[
 L/32+3r\le L/32+3R\le L/16.
\]
Put $n_{\mathrm{in}}=|N_U\cap B(c,L/32)|$. Every center that can
meet $S$ at an allowed radius lies within $L/8+3R\le L/4$ of $c$.
Let $n_{\mathrm{out}}$ count the centers in
$\{z\in N_U:\dist(z,S)\le L/8\}$ outside the inner set.
Lemma~\ref{lem:net-counts} gives
\[
 n_{\mathrm{in}}\ge\frac{L}{\sigma R},\qquad
 \frac{n_{\mathrm{out}}}{n_{\mathrm{in}}}\le\frac{12\sigma}{\eps^2}.
\]

Order the other relevant centers $z_1,\ldots,z_{n_{\mathrm{out}}}$
by nondecreasing $\alpha_i=\dist(z_i,S)$, with fixed tie-breaking.
If $S$ is split by the initial partition, the first center whose
ball meets it cannot cover it:
no point of $S$ has yet been assigned, so a first ball covering $S$
would place it in one part. That center is some $z_i$ for which
\begin{enumerate}
\item $\alpha_i\le\beta<\alpha_i+\delta_C$; smaller radii miss $S$,
      and radii at least $\alpha_i+\delta_C$ cover it by the triangle
      inequality;
\item $z_i$ precedes the $n_{\mathrm{in}}$ inner centers and
      $z_1,\ldots,z_{i-1}$, all of which meet $S$ when
      $\beta\ge\alpha_i$.
\end{enumerate}
The two events depend on independent choices. The first has probability
at most $16\delta_C/L$, since the radius interval has length $L/16$.
In a uniform random order, each of these $n_{\mathrm{in}}+i$
centers is equally likely to come first, so the second event has
probability $1/(n_{\mathrm{in}}+i)$. Thus
\begin{align*}
 \Pr[S\text{ is split}]
 &\le\frac{16\delta_C}{L}
        \sum_{i=1}^{n_{\mathrm{out}}}\frac1{n_{\mathrm{in}}+i}\\
 &\le\frac{16\delta_C}{L}
        \log\left(1+\frac{n_{\mathrm{out}}}{n_{\mathrm{in}}}\right)\\
 &\le\frac{\chi r}{L}\log\left(1+\frac{12\sigma}{\eps^2}\right),
\end{align*}
where $\chi=96$ and $\delta_C\le6r$. If the initial partition preserves $S_C$, the adjustment preserves $C$.
Summing over large clusters proves the expectation bound; no independence
between their splitting events is needed.
\end{proof}

\begin{proof}[Proof of Proposition~\ref{prop:partition}]
We first verify coverage, the diameter bound, and preservation of small clusters.
For any order $\pi$ and radius $\beta\in[L/16,L/8]$, maximality of
$N_U$ places every point of $U$ within $2\widehat R\le\beta$ of a
center. Thus the initial parts cover $U$ and have diameter at most
$2\beta\le L/4$. Since $2\widehat b<\widehat R$, each component
$K$ moved by the routine lies in one $G_{\widehat R}$-component.
Every moved point is within $\diam(K)<2\widehat b$ of its
representative in the receiving initial part. The final regions from
wide components therefore have diameter less than
\[
 L/4+4\widehat b<L.
\]
Components kept whole have diameter at most $L$. For $\widehat R=R$,
Lemma~\ref{lem:small-components} implies that no small optimal
cluster is split. Every optimal cluster is connected through its
center in $G_R$, so each $G_R$-component is a union of whole optimal
clusters.

The radius
list has $O(n^2)$ entries, and each partition takes polynomial time
to compute. If $V=\varnothing$, each $G_R$-component is kept whole
and is a union of optimal clusters at a correct guess, so $Z=0$.

Suppose $V\ne\varnothing$ and $\sigma R\le L\le\rho<2L$.
Since $\rho=\eps M/4<2L$, we have $M/L<8/\eps$.
Lemma~\ref{lem:cut} gives
\[
 \mathbb E Z
 \le\frac{\chi M}{L}\log\left(1+\frac{12\sigma}{\eps^2}\right)
 <\frac{8\chi\Lambda_\eps}{\eps}
 \le\frac{2F_\eps}{3}.
\]
Here $\log x\le\log_2x\le\lceil\log_2x\rceil$ for $x>1$.
Markov's inequality therefore gives
$\Pr_{\pi,\beta}[Z\le F_\eps]\ge1/3$.

For a fixed order, every test involving $\beta$ has the form
$d(z,x)\le\beta$, so the partition is constant on every open
interval $(\beta_i,\beta_{i+1})$. The breakpoints have probability zero.
Whenever an order and a radius outside the breakpoints give
$Z\le F_\eps$, the midpoint of that interval gives the same
partition. Thus the probability, over $\pi$ alone, that the
enumerated family contains such a partition is at least $1/3$.
\end{proof}

\subsection{Local solver implementation}
\label{sec:local-construction}

The routine $\texttt{SolveRegion}(\widehat R,W)$ returns a table
indexed by the number of clusters, or rejects. For nonempty $W$, retain
the candidates
in~\eqref{eq:candidates} contained in $W$, with their original
representations and actual radii as costs. All ball tests use the full
set $P$ and the global parameters $\widehat R$ and
$\widehat b=\eps\widehat R/4$.

A global $G_{2\widehat b}$-component $K$ is \emph{optional in $W$}
if $K\subseteq W$ and $r(K)<\widehat b$. A component only partially
contained in $W$ is not optional. The routine selects disjoint
candidates. Each optional component must be fully covered or fully
uncovered; an uncovered component becomes one cluster. Every point
outside the optional components must be covered.

Greedily choose a maximal strictly $2\widehat R$-separated set
$N_W\subseteq W$, and form a graph $\mathcal H_W$ on $N_W$ with edges
between points at distance at most $8\widehat R$. Assign each candidate
to the vertex nearest its specified center, using fixed tie-breaking.
The variable at each vertex $v\in N_W$ records a disjoint family of
at most $h_\eps$ candidates assigned to $v$, including the empty family. Thus each
variable has $n^{O(1/\eps)}$ possible values.

Impose the following constraints and contributions:
\begin{itemize}
\item Every point is covered at most once, and every nonoptional point
      is covered exactly once.
\item All points of an optional component $K$ have the same coverage
      indicator. If it is zero, $K$ contributes one cluster and cost
      $r(K)$; otherwise it contributes neither.
\item Each vertex contributes the number of selected candidates
      assigned to it and the sum of their actual radii.
\end{itemize}
Appendix~\ref{app:dp} defines these contributions on every assignment,
including assignments violating the constraints.

\medskip\noindent\textbf{Remark (optional components).}
An optional component need not be an optimal small cluster. At the
correct guess it can instead meet several large optimal clusters.
Its points may then be covered by different candidates: the
constraint requires equal coverage indicators, not a common
covering candidate. Selecting all true large clusters covers every
point of such a component exactly once.

We use the following algorithm to obtain a tree decomposition.

\begin{theorem}[Korhonen {\cite[Theorem~I.1]{Korhonen21}}]
\label{thm:treewidth-algorithm}
Given a graph $H$ and an integer $w\ge0$, a deterministic algorithm
runs in time $2^{O(w)}|V(H)|$ and either returns a tree decomposition
of width at most $2w+1$ or correctly reports $\tw(H)>w$.
\end{theorem}

Apply this algorithm to $\mathcal H_W$ with $w=\omega_\eps$, rejecting
only if it reports $\tw(\mathcal H_W)>\omega_\eps$. On a returned
decomposition, run the dynamic program in Appendix~\ref{app:dp}.
Its states record the candidate choices at the vertices in a bag
and a cluster count. Each constraint is checked once, and each
contribution is counted once.
For every $0\le j\le|W|$, return the minimum cost $\DP_W(j)$ and a
corresponding partition, or $+\infty$ if no assignment with count $j$
is available.

\begin{proof}[Proof of Proposition~\ref{prop:local}]
The empty region is immediate, so assume $W\ne\varnothing$.
We first show that the optimum is a permitted assignment, then
verify that the constraints can be enforced by the dynamic program.

On a stable input at $\widehat R=R$, with $W$ a union of optimal
clusters, Lemmas~\ref{lem:separation} and~\ref{lem:small-components}
show that the restricted optimum consists of candidates and
uncovered optional components. For every true large cluster
$C\subseteq W$, select exactly one representation $(c,r(C))$,
choosing an optimal center $c$ by input order.
Every point of $W$ is within $2R$ of $N_W$, so each such cluster
assigned to $v$ lies in $B(v,3R)$. Lemma~\ref{lem:ball-budget}
bounds their total radius by $3R$, even if $v$ lies outside their
union. Hence at most $3R/b=12/\eps$ are assigned to $v$, and
the limit of $h_\eps$ selected candidates per vertex retains the optimum.

On every input, each constraint involves a clique of $\mathcal H_W$.
The specified center $c$ of a retained candidate
$A=B(c,s)\subseteq W$ belongs to $A$ and hence to $W$, so its
assigned vertex is within $2\widehat R$ of $c$.
Since $s\le\widehat R$, a candidate containing $x$ is assigned
to a vertex within $3\widehat R$ of $x$. The vertices involved
in a point constraint are therefore pairwise within $6\widehat R$.
For an optional component $K$, $\diam(K)<2\widehat b$, so its
constraint involves vertices pairwise closer than
$6\widehat R+2\widehat b<8\widehat R$.
Each vertex's own cost and cluster count depend only on its variable.

A tree decomposition is a tree of bags such that every graph vertex
appears, the endpoints of each edge appear together in some bag, and
the bags containing any vertex form a connected subtree. The width
is the largest bag size minus one. Every
clique lies in a bag, so each constraint and contribution can be
assigned to a containing bag. The decomposition returned by
Theorem~\ref{thm:treewidth-algorithm} has width at most
$2\omega_\eps+1=O(\eps^{-4})$. A bag therefore has
$O(\eps^{-4})$ variables, each with $n^{O(1/\eps)}$ values.
Combining tables over cluster counts adds only a polynomial factor,
so the recurrence in Appendix~\ref{app:dp} runs in
$n^{O(\eps^{-5})}$ time. For $n\ge2$, this also absorbs the
decomposition algorithm's $2^{O(\eps^{-4})}$ factor.
These bounds hold on every input, and the constraints ensure that
every finite entry represents a partition of $W$ with its actual cost.

Finally, on a stable input with $\widehat R=R$ and aligned $W$
of diameter at most $\rho$, Lemma~\ref{lem:treewidth} prevents
rejection. The permitted assignment above gives table cost at most
$\cost(\OPT|_W)$ at the true cluster count. Feasibility and
Lemma~\ref{lem:aligned} give equality and the same unique partition.
\end{proof}

\subsection{Full correctness and bit-complexity analysis}
\label{app:full-correctness}

\begin{lemma}
\label{lem:scale-dichotomy}
If $\rho<\sigma R$, then $t\le N_\eps$. Otherwise the list
\[
 L=2^jR,\qquad \log_2\sigma\le j\le\lceil\log_2n\rceil
\]
contains a value satisfying $\sigma R\le L\le\rho<2L$ and
$M/L<8/\eps$. For this value, a call to
$\texttt{RandomPartitions}(R,L)$ returns a family containing a
partition with at most $F_\eps$ split large clusters with probability
at least $1/3$.
\end{lemma}

\begin{proof}
If $\rho<\sigma R$, then $M<4\sigma R/\eps$, so
$t\le M/b<16\sigma/\eps^2\le N_\eps$.
Otherwise choose the largest value $L=2^jR$, with integer $j$, that
does not exceed $\rho$. Since $\sigma=512=2^9$, it satisfies
$\sigma R\le L\le\rho<2L$. Also $M/L<8/\eps$, because
$\rho=\eps M/4$.
We have $\rho\le\eps nR/4<nR$, so this value occurs in the stated
list. In particular, $\rho\ge512R$ implies $n>512$, ensuring that
the index range is nonempty. The probability bound now follows from
Proposition~\ref{prop:partition}.
\end{proof}

\begin{proof}[Proof of Theorem~\ref{thm:randomized-recovery}]
The case $k=n$ is immediate, so suppose $k<n$.
The initialized answer is feasible, and every proposal in Branch A is
checked explicitly. In Branch B, the guessed candidates are disjoint,
the residual regions partition their complement, and every finite
local table entry gives a partition of its region.
Equation~\eqref{eq:combine-regions} enforces exactly $k$ clusters.
Thus every output is feasible, regardless of the sampled permutations.

Now suppose the input is stable and consider the guess $\widehat R=R$.
If $\rho<\sigma R$, Lemma~\ref{lem:scale-dichotomy} gives at most
$N_\eps$ large clusters. Each has a candidate representation
$(c,r(C))$ with $c$ an optimal center, by
Lemma~\ref{lem:separation}, so Branch A enumerates their family.
The remaining small clusters are global $G_{2b}$-components by
Lemma~\ref{lem:small-components}.
Deleting the large clusters does not change these components.
This iteration therefore proposes $\OPT$, for every choice of the
random permutations.

If $\rho\ge\sigma R$, fix the value $L$ supplied by
Lemma~\ref{lem:scale-dichotomy}. With probability at least $1/3$,
the call at $(\widehat R,L)=(R,L)$ returns a family containing a
partition that splits at most $F_\eps$ large clusters. Condition on
this event. Branch B enumerates this partition and the family
$\mathcal X$ of its split large clusters. Every remaining optimal
cluster lies in a single region: no small cluster is split, by
Proposition~\ref{prop:partition}, and every split large cluster has
been removed. Each residual set $W_Q$ is a union of optimal clusters
and has diameter at most $L\le\rho$.
By Proposition~\ref{prop:local}, its local table contains the
restricted optimum at the true count. These counts together with
$|\mathcal X|$ sum to $k$. Combining the tables therefore gives a
feasible partition of cost at most $\cost(\OPT)$, which must be
$\OPT$ by uniqueness. Since the algorithm retains only cheaper
feasible partitions, this event guarantees that its final output is
$\OPT$. Only this one call needs to succeed.

There are at most $n^2$ guesses, $O(\log n)$ values of $L$ per guess,
and $n^2$ candidate representations. Each call to
$\texttt{RandomPartitions}$ samples one permutation and returns
$O(n^2)$ partitions. A uniform permutation of at most $n$ elements
can be generated with $O(n)$ uniform integer choices.
Branch A enumerates $n^{O(N_\eps)}=n^{O(\eps^{-2})}$ families.
Branch B enumerates
$n^{O(F_\eps)}=n^{O(\eps^{-1}\log(2/\eps))}$ families per partition
and solves at most $n$ local problems per family. Each local problem
takes $n^{O(\eps^{-5})}$ time by Proposition~\ref{prop:local}.
Combining tables and checking feasibility add polynomial factors.
The total is therefore
\[
 n^{O(N_\eps+F_\eps+h_\eps(\omega_\eps+1)+1)}
 =n^{O(\eps^{-5})}.
\]
These bounds hold on arbitrary inputs and for every sampled permutation.

For independent runs, the probability that none returns $\OPT$ is at
most $(2/3)^m$. If any run returns $\OPT$, the cheapest output is
$\OPT$, since every output is feasible. This proves the amplified
guarantee and its running-time bound.
\end{proof}

\paragraph{Bit complexity.}
For rational input and a fixed rational parameter $\gamma$, all thresholds and
radius midpoints have polynomial binary length: $L=2^j\widehat R$
with $j=O(\log n)$, and $\widehat b=\eps\widehat R/4$.
Each cluster radius is an input distance, and each stored cost is a
sum of at most $n$ such distances. Thus every arithmetic operation
and comparison takes $\operatorname{poly}(B_{\mathrm{in}})$ bit time.
The order family in Appendix~\ref{app:derandomization} uses integers
of $O(\log n)$ bits, so the deterministic algorithm has the same
polynomial bit overhead.
Rejection sampling implements a uniform integer choice from at most
$n$ options using $O(\log n)$ expected fair bits, giving the
randomized algorithm's expected bit bound.

Recovery does not require $\gamma$ to be rational. For each fixed
real $1<\gamma<2$, choose a fixed rational $\gamma_0$ with
\[
 \frac{\gamma-1}{2}\le\gamma_0-1<\gamma-1.
\]
Every $\gamma$-stable instance is also $\gamma_0$-stable. Using
$\gamma_0$ as the algorithm's parameter therefore gives the same
recovery guarantee, with
$n^{O((\gamma_0-1)^{-5})}=n^{O((\gamma-1)^{-5})}$ running time
and the bit bounds above. For $\gamma\ge2$, the fixed rational
parameter $3/2$ gives polynomial bit time.

\subsection{Dynamic programming recurrence}
\label{app:dp}

We give the recurrence for the dynamic program used by
$\texttt{SolveRegion}$ in Section~\ref{sec:local-construction}.

Let $\mathcal D_v$ be the set of permitted families of candidates
assigned to $v\in N_W$ in Section~\ref{sec:local-construction}, including the
empty family. For a choice of one family at every vertex, let
$m_x$ count the selected candidates containing $x\in W$. Require
$m_x\le1$ everywhere and $m_x=1$ outside optional components. For
each optional component $K$, require all $m_x$, $x\in K$, to be
equal. An uncovered $K$ contributes one cluster and cost $r(K)$;
a covered $K$ contributes neither. Each selected candidate
contributes one cluster and its actual radius. Thus every feasible
assignment specifies a partition with these costs and counts.

Define each contribution for every choice of the variables it depends
on, including choices that violate the constraints: an optional $K$
contributes one cluster if $m_x=0$
for every $x\in K$, and zero otherwise. Its cost is $r(K)$ times
this count. These definitions agree with the preceding contributions on
feasible assignments and allow evaluation before constraints assigned
to other bags have been checked.

The constraint for a point $x$ depends on the variables at vertices
to which candidates containing $x$ are assigned. The constraint for
an optional component $K$ depends on the variables at all such
vertices for $x\in K$. Each of these vertex sets forms a clique in
$\mathcal H_W$ by Section~\ref{sec:local-construction}. Each vertex's cost and cluster
count depend only on its own variable. Root the tree decomposition and
assign every constraint and every contribution to a bag containing
all vertices whose variables it depends on, exactly once; assign
those involving no variables to the root bag.

For a node $a$ with bag $B_a$, an assignment
$\alpha\in\prod_{v\in B_a}\mathcal D_v$, and
$j\in\{0,\ldots,|W|\}$, let $D_a(\alpha,j)$ be the minimum cost
of assignments in the subtree rooted at $a$ that agree with $\alpha$
on $B_a$, satisfy all constraints assigned in that subtree, and have
exactly $j$ clusters counted there. Infeasible states have value
$+\infty$. Let $w_a(\alpha)$ and $q_a(\alpha)$ be the cost and count
assigned to $a$, evaluated after its constraints pass. For a child
$u$ of $a$, put $I_{au}=B_a\cap B_u$ and define
\[
 M_{u\to a}(\eta,\ell)
 =\min_{\substack{\alpha_u\in\prod_{v\in B_u}\mathcal D_v\\
                    \alpha_u|_{I_{au}}=\eta}}
   D_u(\alpha_u,\ell).
\]
For an assignment passing the constraints at $a$, the recurrence is
\[
 D_a(\alpha,j)=w_a(\alpha)+
 \min_{\substack{(j_u)_{u\text{ child of }a}\\
                  \sum_u j_u=j-q_a(\alpha)}}
 \sum_{u\text{ child of }a}
 M_{u\to a}(\alpha|_{I_{au}},j_u).
\]
At a leaf, the value is $w_a(\alpha)$ precisely when
$j=q_a(\alpha)$. At the root $a_0$, set
\[
 \DP_W(j)=\min_{\alpha}D_{a_0}(\alpha,j).
\]

The running-intersection property makes agreement on adjacent bags
equivalent to a consistent choice at every vertex. Every constraint and
contribution is assigned once, so the recurrence counts each cluster
and its cost once. Store minimizing choices and backtrack to recover
the selected candidates and uncovered optional components. Their
point constraints give disjoint coverage of $W$, proving that every
finite entry stores a feasible partition. Counts are
nonnegative, so discarding partial counts above $|W|$ cannot exclude
a feasible partition.

Each variable has $n^{O(1/\eps)}$ values, and each bag has at most
$2\omega_\eps+2$ vertices, since the returned decomposition has
width at most $2\omega_\eps+1$. The count index has at most $n+1$
values. Computing the tables $M_{u\to a}$ and combining cluster counts add
polynomial factors.
The running time is therefore $n^{O((\omega_\eps+1)/\eps)}$, including the
polynomial number of constraints.

\section{Derandomization}
\label{app:derandomization}

We construct the deterministic family of orders used in the proof
of Theorem~\ref{thm:main}. The construction starts from the following
theorem of Indyk.

\begin{theorem}[Indyk {\cite[Theorem~1.1 of the author version]{Indyk01}}]
\label{thm:indyk-strict-minimum}
There are absolute constants $c>1$ and an integer $\ell\ge2$ with
the following property.\footnote{For
absolute constants $c,c'>1$, the theorem states: ``for any
$\epsilon>0$ and $s\le\epsilon n/c$ any $c'\log1/\epsilon$-wise
independent family of functions is $\epsilon$-min-wise independent.''
The source uses $[n]=\{0,\ldots,n-1\}$ for both the domain and range;
its $n$ is our $q$. Its preceding definition concerns $|X|<s$ and
$x\notin X$, with strict-minimum probability
$(1\pm\epsilon)/(|X|+1)$. We set $\epsilon=1/4$,
$s=|T|$, $X=T\setminus\{z\}$, and $x=z$.}
For every positive integer $q$, let
$h:\{0,\ldots,q-1\}\to\{0,\ldots,q-1\}$ be an $\ell$-wise
independent uniform hash. For every nonempty set
$T\subseteq\{0,\ldots,q-1\}$ of size $s\le q/(4c)$ and $z\in T$,
\[
 \Pr[h(z)<h(y)\text{ for all }y\in T\setminus\{z\}]
 \le\frac{5}{4s}.
\]
\end{theorem}

We turn these hash functions into orders, accounting for ties explicitly.

\begin{theorem}[Permutation family]
\label{thm:minwise-family}
For every nonempty finite ordered set $V$, one can deterministically
enumerate, in $|V|^{O(1)}$ time, a nonempty multiset
$\mathcal F_V$ of $|V|^{O(1)}$ permutations of $V$ such that every
nonempty $T\subseteq V$ and $z\in T$ satisfy
\[
 \Pr_{\pi\in\mathcal F_V}
 \bigl[z\text{ precedes every point of }T\setminus\{z\}
       \text{ in }\pi\bigr]
 \le\frac{3}{2|T|}.
\]
The probability is uniform over the multiset, counting repetitions.
\end{theorem}

\begin{proof}
Fix $c,\ell$ from Theorem~\ref{thm:indyk-strict-minimum}; these
constants do not depend on the input.
Put $m=|V|$. Choose a prime $q=O(m^2)$ satisfying
\[
 q\ge\max\{4cm,4m^2\},
\]
and label $V$ by distinct elements of $\mathbb F_q$. Enumerate every
polynomial of degree at most $\ell-1$ over $\mathbb F_q$. For each
polynomial, sort the elements of $V$ by their values, interpreted as
integers in $\{0,\ldots,q-1\}$, breaking ties by the given order on
$V$. Retain the resulting orders with multiplicity to obtain
$\mathcal F_V$.

Uniform coefficients give $\ell$-wise independent uniform hash
values on distinct field elements. Fix $T\subseteq V$ of size
$s\ge1$ and $z\in T$. If $z$ is first in the resulting order, either
its hash is strictly smallest or it shares a hash value with another
point of $T$. Pairwise independence and a union bound give the latter
event probability at most
\[
 (s-1)/q\le1/(4s).
\]
Since $s\le m\le q/(4c)$, Indyk's theorem bounds the former by
$5/(4s)$. Their sum is $3/(2s)$, as required.

There are $q^\ell=m^{O(1)}$ coefficient tuples. A suitable prime can
be found by deterministic search in polynomial time, and evaluating
the constant-degree polynomials and sorting also take polynomial
time. Thus the multiset is nonempty and can be enumerated within the
claimed time bound.
\end{proof}

Theorem~\ref{thm:minwise-family} supplies everything needed from this
appendix in the proof of Theorem~\ref{thm:main}.

\section{Proofs for recognizing stability}
\label{app:recognition-proof}

We prove Theorem~\ref{thm:recognition-hardness}, with rational input
distances and a fixed rational $\gamma>1$. Perturbations may still
be real-valued.

Let $\operatorname{OPT}_j(Y)$ denote the minimum cost of an exact-$j$
partition of a nonempty finite metric $(Y,d_Y)$. Given $1\le k\le |Y|$
and $\tau>0$, deciding whether $\operatorname{OPT}_k(Y)\le \tau$ is NP-hard
even for integer distances of polynomial binary length
\cite[Section~3 of the author version]{Gibson10}.
The ball-cover formulation permits at most $k$ balls centered at input points, including
radius-zero balls. Lemma~\ref{lem:cover-conversion} and the converse
map from clusters to centered balls give equality of the two optimum
values. Splitting off singletons while retaining a center in each remaining
cluster also gives
\[
 \operatorname{OPT}_j(Y)\ge\operatorname{OPT}_k(Y)
 \qquad(1\le j\le k\le |Y|).
\]

The construction has a known optimum of cost $R_0$, while its cheapest
alternative has cost $\delta+\operatorname{OPT}_k(Y)$, with $R_0$ and
$\delta$ defined below. A metric
perturbation raises the known optimum to that alternative's cost
without changing the alternative.

Fix $\gamma>1$ and write
\[
 D=\diam(Y),\qquad
 R_0=1+D+\tau+\frac{\tau}{\gamma-1},\qquad
 \delta=\gamma R_0-\tau,\qquad L_0=\delta+R_0+1.
\]
For a singleton source, $D=0$. These choices ensure
\begin{equation}
\label{eq:recognition-scales}
 R_0>D,\qquad R_0>\tau,\qquad (\gamma-1)R_0>\tau,\qquad
 \delta>R_0,\qquad L_0>\gamma R_0.
\end{equation}
Adjoin a hub $u_0$ to $Y$, put $X=Y\cup\{u_0\}$, and add a disjoint set
$Z$ of $k+1$ dummy points. On $P=X\cup Z$ retain the distances in $Y$
and set
\begin{equation}
\label{eq:recognition-metric}
 \begin{aligned}
 d(u_0,y)&=R_0 &&(y\in Y),\\
 d(z,z')&=\delta &&(z,z'\in Z,\ z\ne z'),\\
 d(x,z)&=L_0 &&(x\in X,\ z\in Z).
 \end{aligned}
\end{equation}
The target cluster count is $K=k+2$, and its supplied partition is
\[
 \mathcal O_0=\{X\}\cup\{\{z\}:z\in Z\}.
\]
The instance has $|Y|+k+2$ points. Its distances have polynomial
binary length, and the construction takes polynomial time.
It is metric: hub/source triangles have sides
$R_0,R_0,d_Y(y,y')$ with $d_Y(y,y')<R_0$; dummy triangles are equilateral;
and triangles meeting $X$ and $Z$ have two sides $L_0$ and a third side
less than $L_0$.

\begin{lemma}
\label{lem:recognition-alternative}
Let $a=\operatorname{OPT}_k(Y)$. Then $a\le D<R_0$,
$\cost_d(\mathcal O_0)=R_0$, and
\[
 \min_{\substack{\mathcal Q\text{ an exact-}K\text{ partition of }P\\
                 \mathcal Q\ne\mathcal O_0}}
       \cost_d(\mathcal Q)=\delta+a.
\]
In particular, $\mathcal O_0$ is the unique original optimum.
\end{lemma}

\begin{proof}
Refining a one-cluster solution gives $a\le D$. The hub serves $X$ at
radius $R_0$, and every source point sees the hub at distance $R_0$, so
$r_d(X)=R_0$.

Consider a different exact-$K$ partition $\mathcal Q$. If a cluster meets
both $X$ and $Z$, each possible center sees a point in the other set at
distance $L_0$. Its cost is therefore at least
$L_0>\delta+R_0>\delta+a$.

Otherwise, let $j$ and $\ell$ count its clusters in $X$ and $Z$.
We have $j+\ell=k+2$ and $1\le\ell\le k+1$.
If $j=1$, then $\ell=k+1$ and $\mathcal Q=\mathcal O_0$.
Hence $2\le j\le k+1$ and $\ell\le k$, so at least one dummy cluster
is nonsingleton and costs $\delta$. If $u_0$ shares its cluster with a
source point, that cluster costs $R_0$, giving a total of at least
$\delta+R_0>\delta+a$. If $u_0$ is a singleton, $Y$ has $j-1\le k$
clusters and costs at least $a$. The total is again at least $\delta+a$.

Equality is attained by an optimal $k$-partition of $Y$, the singleton
$\{u_0\}$, and the single cluster $Z$. This uses exactly $k+2$ clusters.
Finally, $\delta+a>R_0$ by~\eqref{eq:recognition-scales}, proving uniqueness.
\end{proof}

\begin{proposition}
\label{prop:recognition-threshold}
For the constructed instance, put
\[
 \lambda_*=\frac{\delta+a}{R_0}
           =\gamma+\frac{\operatorname{OPT}_k(Y)-\tau}{R_0}.
\]
Under either perturbation convention, $\mathcal O_0$ is
$\lambda$-stable exactly when $1\le\lambda<\lambda_*$.
A metric perturbation creates a tie at $\lambda=\lambda_*$.
\end{proposition}

\begin{proof}
Under any $\lambda$-perturbation, the original hub serves $X$ at radius
at most $\lambda R_0$, so $\cost_{d'}(\mathcal O_0)\le\lambda R_0$.
Every competitor has perturbed cost at least its original cost and
hence at least $\delta+a$, by Lemma~\ref{lem:recognition-alternative}.
Thus $\mathcal O_0$ remains uniquely optimal whenever $\lambda<\lambda_*$.

At $\lambda_*$, change only the hub-to-source distances from $R_0$ to
$\lambda_*R_0=\delta+a$. Every point of $X$ now has eccentricity
$\delta+a$, since the unchanged source distances are at most $D<R_0$.
The partition consisting of an optimal $k$-partition of $Y$,
$\{u_0\}$, and $Z$ uses none of the changed distances and retains cost
$\delta+a$. It therefore ties $\mathcal O_0$.

The perturbation remains metric. Hub/source triangles have two equal
hub sides $\delta+a$ and a source side at most $D$.
Triangles spanning $X$ and $Z$ retain two sides $L_0$ and a third side
less than $L_0$, because $\delta+a<\delta+R_0<L_0$. All other triangles
are unchanged. This same witness is allowed at every
$\lambda\ge\lambda_*$.
\end{proof}

It follows that
\begin{equation}
\label{eq:recognition-reduction}
 \mathcal O_0\text{ is }\gamma\text{-stable}
 \quad\Longleftrightarrow\quad
 \operatorname{OPT}_k(Y)>\tau.
\end{equation}
Since $\mathcal O_0$ is the unique original optimum, recognition and
verification have the same answer. This proves the hardness claims in
Theorem~\ref{thm:recognition-hardness}. Computing $\lambda_*$ also
recovers the source optimum by $a=R_0\lambda_*-\delta$. The reduction
is exact; $\lambda_*-\gamma$ need not be bounded away from zero.

For membership, we use certificates of instability.

\begin{lemma}
\label{lem:instability-certificate}
For rational input distances and rational $\gamma\ge1$, instability
of a specified partition under either perturbation convention has a
rational witness of polynomial binary length and can be verified in
polynomial time.
\end{lemma}

\begin{proof}
Suppose that a partition $\mathcal O$ fails to be uniquely optimal
under an allowed, possibly real-valued perturbation. Choose a distinct
competing partition $\mathcal Q$ and an optimal center $z_B\in B$
for each $B\in\mathcal Q$ under that perturbation. For every center
$c$ in each cluster $C\in\mathcal O$, choose $x_c\in C$ attaining
its eccentricity.

Use symmetric distance variables $p_{uv}=p_{vu}$, radius lower bounds
$t_C$ for the candidate clusters, and radius upper bounds $u_B$ for the
competitor clusters. With $D_{\max}=\gamma\diam(P)$, impose
\[
 \begin{gathered}
 d(u,v)\le p_{uv}\le\gamma d(u,v),\qquad p_{uu}=0,\\
 p_{c x_c}\ge t_C\quad(c\in C,\ C\in\mathcal O),\qquad
 p_{z_Bx}\le u_B\quad(x\in B,\ B\in\mathcal Q),\\
 \sum_{C\in\mathcal O}t_C\ge\sum_{B\in\mathcal Q}u_B,\qquad
 0\le t_C,u_B\le D_{\max}.
 \end{gathered}
\]
For metric perturbations, additionally impose
$p_{uv}\le p_{uw}+p_{wv}$ for every triple.

The original witness makes this system feasible by using its actual
distances and cluster radii. Conversely, every feasible point satisfies
\[
 \cost_p(\mathcal O)\ge\sum_Ct_C
 \ge\sum_Bu_B\ge\cost_p(\mathcal Q).
\]
The first inequality holds because every possible center of $C$
is at distance at least $t_C$ from its chosen point $x_c$; the last holds because
$z_B$ serves all of $B$ at distance at most $u_B$.

This bounded polyhedron has polynomially many variables and constraints,
with rational coefficients of polynomial binary length. If it is
nonempty, it has a vertex whose coordinates also have polynomial
binary length:
after clearing denominators, Cramer's rule and the determinant bound
$|\det A|\le m!H^m$ for an $m\times m$ integer matrix with entries
of absolute value at most $H$ give the required encoding bound.
Thus any real perturbation witnessing instability can be replaced
by a rational one of polynomial binary length. The certificate
consists of this perturbation and the distinct competing partition
$\mathcal Q$. Checking that $\mathcal Q$ is a distinct feasible
partition, checking the perturbation bounds (and triangle inequalities,
when required), and comparing the two partition costs takes polynomial time.

For a singleton candidate cluster, $x_c=c$ forces $t_C=0$.
The case of no distinct feasible partition has no instability witness.
\end{proof}

Lemma~\ref{lem:instability-certificate} puts specified-partition
verification in coNP under either convention. Together with the
reduction, this proves part~(i) of Theorem~\ref{thm:recognition-hardness}.

The recovery algorithm removes the need to supply a candidate partition.

\begin{proposition}
\label{prop:recognition-upper}
For rational input and every fixed rational $\gamma>1$, recognizing
stability under arbitrary symmetric perturbations belongs to coNP.
\end{proposition}

\begin{proof}
Run the deterministic recovery algorithm of
Theorem~\ref{thm:main}, constructed in its proof in
Section~\ref{sec:guarantee},
with fixed tie-breaking, to obtain a feasible candidate partition
in polynomial bit time. An input is unstable if and only if this
candidate admits an instability certificate: stable inputs give
their stable optimum, whereas on unstable inputs every candidate
fails verification. Apply Lemma~\ref{lem:instability-certificate}
to this candidate to obtain a polynomially checkable certificate,
proving coNP membership.
\end{proof}

This proves part~(ii) of Theorem~\ref{thm:recognition-hardness}.
The membership argument uses recovery under arbitrary symmetric
perturbations and does not apply to instances stable only under metric
perturbations. For metric perturbations, recognition is shown only
to be coNP-hard.

\section{A consequence for power costs}
\label{app:radius-costs}

For fixed $0<p\le1$, consider
$\cost^{(p)}_{d'}(\mathcal P)=\sum_{C\in\mathcal P}r_{d'}(C)^p$.
Stability refers to this objective on exact-$k$ partitions with
in-cluster centers, under arbitrary symmetric perturbations.
The distance transformation $d\mapsto d^p$ appears in
Banerjee et al.~\cite[Section~6.1.3 of the full version]{Banerjee25};
its effect on stability gives a direct consequence of our main theorem.

\begin{corollary}[Power costs]
\label{cor:power-recovery}
Fix $0<p\le1$ and $\gamma>1$. For every finite metric $d$,
$\widetilde d=d^p$ is a metric, and
\[
 (P,d,k)\text{ is $\gamma$-stable for }\cost^{(p)}
 \quad\Longleftrightarrow\quad
 (P,\widetilde d,k)\text{ is $\gamma^p$-stable for MSR}.
\]
Under these equivalent stability conditions, the common unique optimal
partition can be recovered deterministically in $n^{O(\eta^{-5})}$ time,
where $\eta=\min\{\gamma^p-1,1\}$.
On every finite metric input with $1\le k\le n$, the algorithm halts
and returns a feasible exact-$k$ partition.
\end{corollary}

\begin{proof}
The inequality $(a+b)^p\le a^p+b^p$ for $a,b\ge0$ shows that
$d^p$ satisfies the triangle inequality. Since $t\mapsto t^p$ is
strictly increasing, finite minima and maxima commute with it, giving
$r_{(d')^p}(C)=r_{d'}(C)^p$ for every symmetric perturbation $d'$.
Raising each distance to the power $p$ gives a bijection between the
perturbations $d\le d'\le\gamma d$ and those satisfying
$d^p\le\widetilde d'\le\gamma^p d^p$; the inverse is exponentiation
by $1/p$. Thus partition costs agree under corresponding
perturbations, proving the stability equivalence. Compute $d^p$
and apply Theorem~\ref{thm:main} with stability factor $\gamma^p$.
\end{proof}

The running time counts exact arithmetic and power evaluations.
The transformed distances may be irrational even for rational input,
so no bit-complexity bound is asserted here.


\begin{thebibliography}{99}
\raggedright
\bibitem{AngelidakisAABCD19}
Haris Angelidakis, Pranjal Awasthi, Avrim Blum, Vaggos Chatziafratis,
and Chen Dan.
\newblock Bilu--Linial stability, certified algorithms and the independent
set problem.
\newblock In \emph{27th Annual European Symposium on Algorithms (ESA)},
volume 144 of LIPIcs, pages 7:1--7:16, 2019.

\bibitem{AngelidakisMM17}
Haris Angelidakis, Konstantin Makarychev, and Yury Makarychev.
\newblock Algorithms for stable and perturbation-resilient problems.
\newblock In \emph{49th Annual ACM SIGACT Symposium on Theory
of Computing (STOC)}, pages 438--451, 2017.

\bibitem{AwasthiBS12}
Pranjal Awasthi, Avrim Blum, and Or Sheffet.
\newblock Center-based clustering under perturbation stability.
\newblock \emph{Information Processing Letters}, 112(1--2):49--54, 2012.

\bibitem{Baker94}
Brenda~S. Baker.
\newblock Approximation algorithms for NP-complete problems on planar graphs.
\newblock \emph{Journal of the ACM}, 41(1):153--180, 1994.

\bibitem{BalcanHW16}
Maria-Florina Balcan, Nika Haghtalab, and Colin White.
\newblock $k$-center clustering under perturbation resilience.
\newblock In \emph{43rd International Colloquium on Automata,
Languages, and Programming (ICALP)}, volume 55 of LIPIcs,
pages 68:1--68:14, 2016.

\bibitem{BalcanLiang16}
Maria-Florina Balcan and Yingyu Liang.
\newblock Clustering under perturbation resilience.
\newblock \emph{SIAM Journal on Computing}, 45(1):102--155, 2016.

\bibitem{Banerjee25}
Sandip Banerjee, Yair Bartal, Lee-Ad Gottlieb, and Alon Hovav.
\newblock Improved fixed-parameter bounds for min-sum-radii and diameters
$k$-clustering and their fair variants.
\newblock In \emph{Proceedings of the AAAI Conference on Artificial
Intelligence}, 39(15):15481--15488, 2025.
\newblock Full version: arXiv:2501.17708.

\bibitem{BenDavidReyzin14}
Shalev Ben-David and Lev Reyzin.
\newblock Data stability in clustering: A closer look.
\newblock \emph{Theoretical Computer Science}, 558:51--61, 2014.

\bibitem{BiluDLS13}
Yonatan Bilu, Amit Daniely, Nati Linial, and Michael Saks.
\newblock On the practically interesting instances of MAXCUT.
\newblock In \emph{30th International Symposium on Theoretical Aspects
of Computer Science (STACS)}, volume 20 of LIPIcs, pages 526--537, 2013.

\bibitem{BiluLinial12}
Yonatan Bilu and Nathan Linial.
\newblock Are stable instances easy?
\newblock \emph{Combinatorics, Probability and Computing},
21(5):643--660, 2012.

\bibitem{BuchemERW24}
Moritz Buchem, Katja Ettmayr, Hugo~K.~K. Rosado, and Andreas Wiese.
\newblock A $(3+\varepsilon)$-approximation algorithm for the minimum
sum of radii problem with outliers and extensions for generalized
lower bounds.
\newblock In \emph{Proceedings of the 2024 Annual ACM-SIAM Symposium
on Discrete Algorithms (SODA)}, pages 1738--1765, 2024.

\bibitem{Campbell23}
Rutger Campbell, Marc Distel, J.~Pascal Gollin, Daniel~J. Harvey,
Kevin Hendrey, Robert Hickingbotham, Bojan Mohar, and David~R. Wood.
\newblock Graphs of linear growth have bounded treewidth.
\newblock \emph{The Electronic Journal of Combinatorics}, 30(3):P3.1, 2023.

\bibitem{DrexlerHKS23}
Lukas Drexler, Jan H\"ockendorff, Joshua K\"onen, and Kevin Schewior.
\newblock Clustering graphs of bounded treewidth to minimize the sum of
radius-dependent costs.
\newblock arXiv:2310.02130, 2023.

\bibitem{FernandezKenyon01}
Wenceslas Fernandez de la Vega and Claire Kenyon.
\newblock A randomized approximation scheme for metric MAX-CUT.
\newblock \emph{Journal of Computer and System Sciences},
63(4):531--541, 2001.

\bibitem{FernandezKKR03}
Wenceslas Fernandez de la Vega, Marek Karpinski, Claire Kenyon, and
Yuval Rabani.
\newblock Approximation schemes for clustering problems.
\newblock In \emph{35th Annual ACM Symposium on Theory of Computing
(STOC)}, pages 50--58, 2003.

\bibitem{FriggstadKS25}
Zachary Friggstad, Kamyar Khodamoradi, and Mohammad~R. Salavatipour.
\newblock Exact algorithms and lower bounds for stable instances of
Euclidean $k$-means.
\newblock \emph{SIAM Journal on Computing}, 54(6):1456--1488, 2025.
\newblock Preliminary version in SODA 2019.

\bibitem{FriggstadRS19}
Zachary Friggstad, Mohsen Rezapour, and Mohammad~R. Salavatipour.
\newblock Local search yields a PTAS for $k$-means in doubling metrics.
\newblock \emph{SIAM Journal on Computing}, 48(2):452--480, 2019.

\bibitem{GaoReyzin25}
Xing Gao and Lev Reyzin.
\newblock Non-center-based clustering under Bilu--Linial stability.
\newblock In \emph{IEEE International Symposium on Information Theory
(ISIT)}, pages 1--5, 2025.

\bibitem{Gibson10}
Matt Gibson, Gaurav Kanade, Erik Krohn, Imran~A. Pirwani, and Kasturi
Varadarajan.
\newblock On metric clustering to minimize the sum of radii.
\newblock \emph{Algorithmica}, 57(3):484--498, 2010.

\bibitem{Gibson12}
Matt Gibson, Gaurav Kanade, Erik Krohn, Imran~A. Pirwani, and Kasturi
Varadarajan.
\newblock On clustering to minimize the sum of radii.
\newblock \emph{SIAM Journal on Computing}, 41(1):47--60, 2012.

\bibitem{GrandoniGY26}
Fabrizio Grandoni, Anupam Gupta, and Jatin Yadav.
\newblock FPT approximation schemes for min-sum radii and min-sum
diameters.
\newblock In \emph{67th IEEE Symposium on Foundations of Computer
Science (FOCS)}, 2026. To appear.
\newblock Full version:
arXiv:2605.10895.

\bibitem{Indyk01}
Piotr Indyk.
\newblock A small approximately min-wise independent family of hash functions.
\newblock \emph{Journal of Algorithms}, 38(1):84--90, 2001.

\bibitem{Korhonen21}
Tuukka Korhonen.
\newblock A single-exponential time 2-approximation algorithm for treewidth.
\newblock In \emph{62nd IEEE Annual Symposium on Foundations of Computer
Science (FOCS)}, pages 184--192, 2021.

\bibitem{KumarMOS26}
Pankaj Kumar, Haiko M\"uller, Sebastian Ordyniak, and Melanie Schmidt.
\newblock On the parameterized complexity of min-sum-radii.
\newblock In \emph{20th Scandinavian Symposium on
Algorithm Theory (SWAT)}, volume 370 of LIPIcs, pages 26:1--26:18, 2026.

\end{thebibliography}
\end{document}